\documentclass[a4paper, 11pt]{article}
\usepackage{amsmath, amssymb, color} 
\input{colordvi.tex}
\usepackage{mathrsfs}
\usepackage{epsfig}
\usepackage[hypertex]{hyperref}
\usepackage[nosort]{cite} 
\def\hsymbu#1{\smash{\lower1.7ex\hbox{\huge$#1$}}}
\newcommand{\bs}{\begin{subequations}}
\newcommand{\es}{\end{subequations}}
\newcommand{\mrm}{\mathrm}
\newcommand{\p}{\partial}
\newcommand{\diff}{\mathrm{d}}
\newcommand{\iu}{\mathrm{i}}
\newcommand{\e}{{\hspace{.1em}{\mathrm e}}}

\newcommand{\Xil}{\Xi_\lambda} 
\newcommand{\Xl}{\mathcal{X}_\lambda} 
\newcommand{\LQ}{\epsilon_Q} 
\newcommand{\Leta}{\epsilon_\eta}
\newcommand{\Ginv}{G^{-1}}
\newcommand{\SW}{S^\mathrm{W}}
\newcommand{\SB}{S^\mathrm{B}}
\newcommand{\PsiW}{\Psi^\mathrm{W}}
\newcommand{\hatG}{\hat{G}}
\newcommand{\hatGinv}{\hat{G}^{-1}}
\newcommand{\xz}{\xi_0}
\newcommand{\ez}{\eta_0}
\newcommand{\calX}{\mathcal{X}}
\newcommand{\calO}{\mathcal{O}}
\newcommand{\hs}{\hspace{.07em}}
\newcommand{\lc}{z} 
\newcommand{\dc}{x} 
\newcommand{\llangle}{\langle\hspace{-0.8mm}\langle}
\newcommand{\bllangle}{\big\langle\hspace{-0.9mm}\big\langle}
\newcommand{\Bllangle}{\Big\langle\hspace{-1.5mm}\Big\langle}

\newcommand{\rrangle}{\rangle\hspace{-0.8mm}\rangle}
\newcommand{\brrangle}{\big\rangle\hspace{-0.9mm}\big\rangle}
\newcommand{\Brrangle}{\Big\rangle\hspace{-1.5mm}\Big\rangle}

\allowdisplaybreaks
\begin{document}
\begin{titlepage}
\rightline{}
\rightline{\tt arXiv:1312.1677}
\rightline{\tt RIKEN-MP-77}
\rightline{\tt UT-Komaba/13-17}
\begin{center}
\vskip 1.5cm
{\Large \bf {
From the Berkovits formulation to the Witten formulation
}}\\
\vskip 0.5cm
{\Large \bf {in open superstring field theory}}\\
\vskip 1.0cm
{\large Yuki {\sc Iimori}$^1$, Toshifumi {\sc Noumi}$^2$, Yuji {\sc Okawa}$^3$ and Shingo {\sc Torii}$^2$}
\vskip 1cm
{\large 
$^1${\it {Department of Physics, Nagoya University, Nagoya 464-8602, Japan}}\\[3mm]
$^2${\it {Mathematical Physics Laboratory, RIKEN Nishina Center, Saitama 351-0198, Japan}}\\[3mm]
$^3${\it {Institute of Physics,  The University of Tokyo,
Komaba, Meguro-ku, Tokyo 153-8902, Japan}}\\[5mm]
iimori@eken.phys.nagoya-u.ac.jp, toshifumi.noumi@riken.jp,\\[2mm]
 okawa@hep1.c.u-tokyo.ac.jp, shingo.torii@riken.jp
}
\vskip 1.5cm
{\large {\bf Abstract}}
\end{center}
\vskip 0.5cm
\baselineskip 16pt

The Berkovits formulation of open superstring field theory
is based on the large Hilbert space of the superconformal ghost sector.
We discuss its relation to the Witten formulation
based on the small Hilbert space.
We introduce a one-parameter family of conditions
for partial gauge fixing of the Berkovits formulation
such that the cubic interaction of the theory under the partial gauge fixing
reduces to that of the Witten formulation in a singular limit.
The local picture-changing operator at the open-string midpoint
in the Witten formulation is regularized in our approach,
and the divergence in on-shell four-point amplitudes
coming from collision of picture-changing operators
is resolved.
The quartic interaction inherited from the Berkovits formulation
plays a role of adjusting different behaviors
of the picture-changing operators
in the $s$ channel and in the $t$ channel
of Feynman diagrams with two cubic vertices,
and correct amplitudes in the world-sheet theory are reproduced.
While gauge invariance at the second order in the coupling constant
is obscured in the Witten formulation
by collision of picture-changing operators,
it is well defined in our approach
and is recovered by including the quartic interaction
inherited from the Berkovits formulation.

\end{titlepage}
\baselineskip 17pt
\tableofcontents

\newpage
\section{Introduction}
\label{section:introduction}
\setcounter{equation}{0}

Gauge invariance plays a fundamental role in the current formulation of covariant string field theory.
The equation of motion of the free theory corresponds to the physical state condition in the world-sheet theory.  
In open bosonic string field theory, for example, it is given by
\begin{equation}
Q \Psi = 0 \,,
\end{equation}
where $\Psi$ is the string field and $Q$ is the BRST operator.
The equivalence relation of the physical states,
\begin{equation}
\Psi \sim \Psi +Q \Lambda \,,
\end{equation}
is implemented as 
the
gauge symmetry
\begin{equation}
\delta \Psi = Q \Lambda
\label{free-gauge-transformation} 
\end{equation}
in string field theory.
In the interacting theory, this gauge symmetry is nonlinearly extended,
and the action is invariant under the nonlinearly extended gauge transformation.
The resulting theory therefore reproduces the spectrum correctly,
and unphysical degrees of freedom do not show up
because of the nonlinearly extended gauge invariance
just as in the Yang-Mills theory.

In open bosonic string field theory~\cite{Witten:bosonic},   
the gauge transformation~(\ref{free-gauge-transformation})  
is extended to
\begin{equation}
\delta \Psi = Q \Lambda +g \, ( \, \Psi \ast \Lambda -\Lambda \ast \Psi \, ) \,,
\label{nonlinearly-extended-gauge-transformation}
\end{equation}
where products of string fields are defined by Witten's star product
and $g$ is the open string coupling constant,  
and the cubic Chern-Simons-like action given by
\begin{equation}
S = -\frac{1}{2} \, \langle \, \Psi, Q \Psi \, \rangle
-\frac{g}{3} \, \langle \, \Psi, \Psi \ast \Psi \, \rangle
\end{equation}
is invariant
under the gauge transformation~(\ref{nonlinearly-extended-gauge-transformation}),
where $\langle \mskip 1.5mu A, B \mskip 2mu \rangle$ is the 
Belavin-Polyakov-Zamolodchikov (BPZ) inner product~\cite{BPZ}
of string fields $A$ and $B$.
While this cubic action is practically useful,
gauge-invariant actions can also be constructed
with higher-order interaction terms
if such interaction terms satisfy a set of relations called
the $A_\infty$ structure~\cite{Stasheff:I, Stasheff:II, G-J, Markl, P-S, Gaberdiel:1997ia}. 
The $A_\infty$ structure is closely related to the covering 
of the moduli space of Riemann surfaces,
and this is the structure underlying the gauge invariance of open bosonic string field theory. 
In closed bosonic string field theory~\cite{Zwiebach:1992ie}, 
the corresponding structure underlying its gauge invariance is called $L_\infty$~\cite{Zwiebach:1992ie, Lada:1992wc, S-S}.

Since the construction of an analytic solution by Schnabl~\cite{Schnabl:2005gv},
powerful analytic methods have been developed in open bosonic string field theory~\cite{Okawa:2006vm,Ellwood:2006ba,Okawa:2006sn,Erler:2006hw,Erler:2006ww,Schnabl:2007az,Kiermaier:2007ba,Fuchs:2007yy,Kishimoto:2007bb,Kiermaier:2007vu,Kiermaier:2008qu,Erler:2009uj,Bonora:2010hi,Kiermaier:2010cf,Murata:2011ex,Bonora:2011ri,Erler:2011tc,Bonora:2011ru,Takahashi:2011wk,Hata:2011ke,Murata:2011ep,Masuda:2012kt,Baba:2012cs,Masuda:2012cj,Bonora:2013cya},
and such analytic methods have been extended
to open superstring field theory~\cite{Erler:2007rh,Okawa:2007ri,Okawa:2007it,Fuchs:2007gw,Erler:2007xt,Kiermaier:2007ki,Aref'eva:2008ad,Gorbachev:2010zz,Erler:2010pr,Noumi:2011kn,Erler:2013wda}.
In open superstring field theory~\cite{Witten:super,Preitschopf:1989fc,Arefeva:1989cp,Berkovits:NS},
we expect that the structure underlying its gauge invariance
be a supersymmetric extension of the $A_\infty$ structure,
and it would be closely related to the covering of the supermoduli space of super-Riemann surfaces. 
However, such an understanding of the gauge invariance
has been developed very little,
and some of the problems we are confronted with in open superstring field theory
seem to be related to the lack of our understanding in this perspective.

For example, in the Witten formulation of open superstring field theory~\cite{Witten:super},
the gauge variation of the action has turned out to be singular
because of the collision of picture-changing operators~\cite{Wendt}. 
There are related divergences in tree-level amplitudes
again caused by the collision of picture-changing operators.
If we recall that the origin of the local picture-changing operator
is the delta-functional support in the gauge fixing
of the world-sheet gravitino field,
it is possible that the source of these divergences is related
to the singular covering of the supermoduli space
of super-Riemann surfaces.
At the moment, however, such an understanding is missing.

On the other hand, the gauge invariance does not
suffer from any singularity in the Berkovits formulation
of open superstring field theory~\cite{Berkovits:NS} in the Neveu-Schwarz (NS) sector.
We do not, however, understand why it works well
in the context of the covering of the supermoduli space
of super-Riemann surfaces.
In the Berkovits formulation, the action contains
interaction vertices whose orders are higher than cubic.
We know that the bosonic moduli space of Riemann surfaces is covered
by Feynman diagrams with cubic vertices alone,
and the higher-order vertices do not contribute
to the covering of the bosonic moduli space.
Since gauge invariance requires the higher-order vertices,
it is expected that the higher-order vertices 
play a role in the covering of the supermoduli space.
At the moment, however, such an understanding is missing.

One possible approach to incorporating the Ramond sector
into the Berkovits formulation was proposed in~\cite{Berkovits:Ramond}.
Compared with the description of the Ramond sector
in the Witten formulation,
this approach is considerably complicated.
In addition, it does not completely respect
the covariance in ten dimensions, although it respects
the covariance for a class of interesting backgrounds
such as D3-branes in the flat ten-dimensional spacetime.
If we understand the relation between the gauge invariance
in the Berkovits formulation and the covering of the supermoduli space
of super-Riemann surfaces, it may lead us to a new approach
to incorporating the Ramond sector.\footnote{
For another attempt to formulate open superstring field theory,
see~\cite{Kroyter:2009rn}.}
Another issue with the Berkovits formulation is that
it has turned out to be formidably complicated
to construct the master action in the Batalin-Vilkovisky formalism~\cite{BV:irreducible, BV:reducible} 
for quantization.
See~\cite{Torii:proceedings, Berkovits:2012}.
In the construction of the master action
in open bosonic string field theory and in closed bosonic string field theory,
the $A_\infty$ structure and the $L_\infty$ structure, respectively,
played a crucial role.
The difficulty in the construction of the master action
in open superstring field theory might also be related
to the lack of our understanding for a supersymmetric extension
of the $A_\infty$ structure.

We therefore think that to explore the relation
between gauge invariance in open superstring field theory
and the covering of the supermoduli space of super-Riemann surfaces
is an important problem for various aspects.\footnote{
See \cite{Jurco:2013qra,Matsunaga:2013mba} for related discussions in closed superstring field theory.} 
In view of recent developments in understanding
the supermoduli space~\cite{Witten:2012ga, Witten:2012bh, Witten:2013cia, Donagi:2013dua}, 
it can be crucially important for the profound question
of whether or not open superstring field theory can be
a consistent quantum theory by itself.
In this paper, as a first step towards this direction,
we address the question of how the divergences
in the Witten formulation can be resolved in the Berkovits formulation.
The Hilbert space of the string field in the Berkovits formulation
is larger than that in the Witten formulation
and, correspondingly, the gauge symmetry in the Berkovits formulation
is larger than that in the Witten formulation.
We perform partial gauge fixing in the Berkovits formulation
to relate it to the Witten formulation.
We introduce a one-parameter family of judicious gauge choices labeled by $\lambda$,
and the cubic interaction in the Berkovits formulation reduces
to that in the Witten formulation in the singular limit $\lambda \to 0$.
We can think of the Berkovits formulation which is partially gauge fixed
with finite $\lambda$ as a regularization of the Witten formulation.
We find that the divergence in the four-point amplitude as $\lambda \to 0$
is canceled by the quartic interaction.
We also find that the divergence in the gauge variation of the action
to the second order in the coupling constant as $\lambda \to 0$
is resolved by incorporating the quartic interaction.
Our approach based on the one-parameter family of gauge choices enables
us to discuss the nature of these divergences in a concrete and well-defined setting,
and it is the main point of this paper.
While higher-point amplitudes in the Berkovits formulation
have not been calculated explicitly,
we do not foresee any singularities.
We thus expect that the theory obtained by the partial gauge fixing
of the Berkovits formulation with finite $\lambda$
provide a consistent formulation in the small Hilbert space,
although further calculations are necessary
to see if it reproduces the amplitudes in the world-sheet theory correctly.
Our next step will be to translate the mechanism of canceling the divergences
discussed in this paper
into the language of the covering of the supermoduli space of super-Riemann surfaces,
and our ultimate goal is to reveal a supersymmetric extension of the $A_\infty$ structure
underlying open superstring field theory.

The organization of the rest of the paper is as follows.
In section~\ref{section:SSFTs}, we briefly review the Witten formulation
and the Berkovits formulation of open superstring field theory.
In section~\ref{section:Gauge-fixing}, we explain the partial gauge fixing
of the Berkovits formulation.
We then use the theory under the partial gauge fixing
as a regularization of the Witten formulation
and discuss the divergence in on-shell four-point amplitudes at the tree level
in section~\ref{section:regularization}
and the divergence in the gauge variation of the action
at the second order in the coupling constant in section~\ref{sec:relation}.
Section 6 is devoted to discussion. 

\section{The Witten formulation and the Berkovits formulation} 
\label{section:SSFTs}
\setcounter{equation}{0}
In this section, we review the Witten formulation~\cite{Witten:super}
and the Berkovits formulation~\cite{Berkovits:NS}
of open superstring field theory, concentrating on 
the NS sector.
The Witten formulation is based on the small Hilbert space
of the superconformal ghost sector
and has the Chern-Simons-like action.
The Berkovits formulation is based on the large Hilbert space
and has the Wess-Zumino-Witten-like (WZW-like) action.
We first summarize the basics of the description of the superconformal ghost sector
in the Ramond-Neveu-Schwarz (RNS) formalism
in subsection~\ref{subsec:small/large}
and briefly review the Witten formulation
in subsection~\ref{Witten formulation}
and the Berkovits formulation
in subsection~\ref{Berkovits formulation}.

\subsection{The superconformal ghost sector}
\label{subsec:small/large}
The superconformal ghost sector in the RNS formalism
can be described in terms of $\xi$, $\eta$, and $\phi$~\cite{FMS1, FMS2,Polchinski:2},
where $\xi$ and $\eta$ are fermionic and $\phi$ is bosonic.
Their fundamental operator product expansions (OPEs) are given by
\begin{equation}
\xi(z_1) \, \eta(z_2) \sim \frac{1}{z_1-z_2}\,,\qquad
\phi(z_1) \, \phi(z_2) \sim -\ln \left(z_1-z_2\right) \,.
\end{equation}
The BRST current $j_B$ in this description
takes the form
\begin{equation} \label{j_B}
j_B
= c\hs T^\mrm{m} +\eta\e^\phi\hs G^\mrm{m} +bc\p c +\frac{3}{4}\hs\p c\p\phi -\frac{1}{4}\hs c\p^2\phi
-\frac{1}{2}\hs c\p\phi\p\phi -c\eta\p \xi -b\eta\p\eta \e^{2\phi} +\frac{3}{4}\hs\p^2 c \,,
\end{equation}
where $T^\mrm{m}$ and $G^\mrm{m}$ are the energy-momentum tensor and the supercurrent, respectively, in the matter sector
and the last term, which is a total derivative, makes the current primary.
Here and in what follows we omit
the normal-ordering symbol
with respect to the SL$(2, \mathbb{R})$-invariant vacuum for simplicity.
The BRST operator is given by
\begin{equation}
Q := \oint_C \frac{\diff z}{2\pi\iu} \, j_B (z) \,.
\end{equation}
We use the doubling trick, and the contour $C$ is along the counterclockwise unit circle 
centered at the origin.

The two important quantum numbers in the open superstring,
the world-sheet ghost number and the picture number,
are defined by the world-sheet ghost number charge 
$Q_{\boldsymbol g}$
and the picture number charge 
$Q_{\boldsymbol p}$,
respectively,  given by
\begin{equation}
Q_{\boldsymbol g} = \oint_C \frac{\diff z}{2\pi\iu} \bigl(  {}-bc(z)-\xi\eta(z)\bigr) \,, \qquad
Q_{\boldsymbol p} = \oint_C \frac{\diff z}{2\pi\iu} \bigl(  {}-\p\phi(z) +\xi\eta (z)\bigr) \,.
\label{Q^gh-Q^pic}
\end{equation}
We summarize the ghost number $\boldsymbol{g}$
and the picture number $\boldsymbol{p}$ 
together with the conformal weight $h$
of various operators in table~\ref{table}.
\begin{table}
\begin{center}
\caption{The ghost number $\boldsymbol{g}$, the picture number $\boldsymbol{p}$, and the conformal weight $h$ of various operators.}
\ \\[-.5ex]
{\renewcommand\arraystretch{1.3}
\begin{tabular}{|c||c|c|c|c|c|c|c|c|}
\hline
operator & $b$ & $c$ & $\xi$ & $\eta$ & $\e^{l\phi}$ & $\beta$ & $\gamma$ & $j_B$ \\
\hline
$(\boldsymbol{g}, \boldsymbol{p})$ & $(-1,0)$ & $(1,0)$ & $(-1,1)$ & $(1,-1)$ & $(0,l)$ & $(-1,0)$ & $(1,0)$ & $(1,0)$ \\
\hline
$h$ & $2$ & $-1$ & $0$ & $1$ & $-\frac{1}{2}\hs l(l+2)$ & $\frac{3}{2}$ & $-\frac{1}{2}$ & $1$ \\
\hline
\end{tabular}
}
\label{table}
\end{center}
\end{table}
Correlation functions on a disk vanish unless the total ghost number is $2$
and the total picture number is $-1$,
with the basic nonvanishing correlation function being
\begin{equation} \label{basic_correlator} 
\langle \, \xi (z) \, c \partial c \partial^2 c (w) \, \e^{-2 \phi (y)} \, \rangle \ne 0 \,. 
\end{equation}
The picture-changing operator $X$ is expressed as~\cite{FMS1, FMS2,Polchinski:2}, 
\begin{equation} \label{PCO}
X := Q \cdot \xi
= \e^\phi G^\mrm{m} +c\p\xi + b\p\eta\e^{2\phi} +\p \bigl( b\eta\e^{2\phi}\bigr)\,.
\end{equation}
The OPE of the picture-changing operator with itself is
\begin{align} \label{XX}
X(z_1) \, X(z_2) &= \bigl\{Q, \xi(z_1)\bigr\} \, X(z_2)
= \bigl\{Q,\, \xi(z_1) \, X(z_2)\bigr\} \nonumber \\*
&\sim
{}-\frac{2}{(z_1-z_2)^2} \, \bigl\{ Q, b\e^{2\phi}(z_2)\bigr\}
-\frac{1}{z_1 - z_2} \, \bigl\{ Q, \p (b\e^{2\phi})(z_2)\bigr\}\,.
\end{align}
As we will discuss later,
the singularity in this OPE causes divergences in the Witten formulation
of open superstring field theory.

The Hilbert space we usually use for the superconformal ghost sector
in the description in terms of the $\beta \gamma$ ghosts
is smaller than the Hilbert space
for the system of $\xi$, $\eta$, and $\phi$
and is called the {\it small Hilbert space}.
Correspondingly, the Hilbert space for the system of $\xi$, $\eta$, and $\phi$ 
is called the {\it large Hilbert space}.
A state in the small Hilbert space
corresponds to a state annihilated by the zero mode of $\eta$
in the description in terms of $\xi$, $\eta$, and $\phi$:
\begin{equation}
\eta_0 A = 0 \,.
\end{equation}
Here and in what follows
we expand an operator $\calO$ of conformal weight $h$ in the coordinate $z$
on the upper half-plane as
\begin{align}
\calO(z) =
\sum_n
\frac{\calO_n}{z^{n+h}}\,.
\end{align}
It follows from the OPEs of $\xi$ and $\eta$
that the zero modes $\xi_0$ and $\eta_0$ satisfy
\begin{equation} \label{xzez}
\xz^2 = \ez^2 = 0\,,\quad \{\xz, \ez\} =1\,.
\end{equation}
Any state $\varphi$ in the large Hilbert space
can be written in terms of two states $A$ and $B$ in the small Hilbert space as
\begin{equation} \label{varphi:decomposition}
\varphi = A +\xi_0 B \,,
\end{equation}
where
\begin{equation}
A = \eta_0 \xi_0 \varphi \,, \qquad
B = \eta_0 \varphi \,.
\end{equation}
We could say that the large Hilbert space is twice as large as the small Hilbert space.

We denote 
the BPZ inner product~\cite{BPZ}
of a pair of states $\varphi_1$ and $\varphi_2$
in the large Hilbert space by $\langle \, \varphi_1, \varphi_2 \, \rangle$.
It vanishes unless the sum of the ghost numbers of $\varphi_1$ and $\varphi_2$ is $2$
and the sum of the picture numbers of $\varphi_1$ and $\varphi_2$ is $-1$.
For any pair of states $A$ and $B$ in the small Hilbert space,
we define the BPZ inner product in the small Hilbert space
$\llangle \, A, B \, \rrangle$ by\footnote{
See appendix B of~\cite{Torii:validity}
for the reason why an imaginary unit is
necessary in the relation between the BPZ inner product in the small Hilbert space
and that in the large Hilbert space.
}
\begin{equation} \label{small/large:BPZ}
\llangle \, A,B \, \rrangle = \iu\hs\langle \, \xz A,\hs B \, \rangle 
= \iu\hs (-1)^A \hs\langle \, A,\hs \xz B \, \rangle \,.
\end{equation}
Here and in what follows a state in the exponent of $-1$ represents
its Grassmann parity: it is $0$ mod $2$ for a Grassmann-even state
and $1$ mod $2$ for a Grassmann-odd state.
The BPZ inner product $\llangle \, A, B \, \rrangle$ vanishes
unless the sum of the ghost numbers of $A$ and $B$ is $3$
and the sum of the picture numbers of $A$ and $B$ is $-2$.

We say that an operator $\mathcal{O} (t)$ is in the small Hilbert space
when $\ez \cdot \mathcal{O} (t) = 0$.
For operators $\mathcal{O}_i$ in the small Hilbert space,
the relation between
correlation functions in the small Hilbert space
denoted by
$\bllangle \, \mathcal{O}_1 (t_1) \, \mathcal{O}_2 (t_2) \, \ldots \mathcal{O}_n (t_n) \,
\brrangle$
and those in the large Hilbert space is given by
\begin{equation}
\label{small-large-correlation-functions}
\bllangle \, \mathcal{O}_1 (t_1) \, \mathcal{O}_2 (t_2) \, \ldots \mathcal{O}_n (t_n) \,
\brrangle
= \iu \, \big\langle \, \xi (t) \, \mathcal{O}_1 (t_1) \, \mathcal{O}_2 (t_2) \, \ldots
\mathcal{O}_n (t_n) \,
\big\rangle \,.
\end{equation}

\subsection{The Witten formulation}
\label{Witten formulation}
The Witten formulation of open superstring field theory~\cite{Witten:super} 
is based on the small Hilbert space,
and it is a natural extension of open bosonic string field theory~\cite{Witten:bosonic}.
The picture number of the string field in the NS sector is $-1$
and that of the string field in the Ramond sector is $-1/2$.
These picture numbers are natural in the context of the state-operator correspondence.
It turned out, however, that the picture-changing operator
inserted at the string midpoint causes divergences
in scattering amplitudes
and in the gauge variation of the action,
and the primary purpose of this paper is to understand
the nature of such divergences.

The action in the Witten formulation is given by
\begin{equation}
\label{Witten:NS-action}
\SW 
= -\frac{1}{2} \, \bllangle \PsiW , Q\PsiW \brrangle 
-\frac{g}{3} \, \bllangle \PsiW,\, X_{\rm mid} \, ( \PsiW\ast\PsiW ) \brrangle \,.
\end{equation}
Here $g$ is the open string coupling constant,
and $\PsiW$ is the open superstring field in the NS sector, which is 
Grassmann odd.\footnote{
We consider a general superconformal field theory in the matter sector
and an appropriate projection analogous to the Gliozzi-Scherk-Olive (GSO) projection
in the case of the flat spacetime in ten dimensions is assumed.}
For later convenience, we have appended the superscript ``W'' to the string field in the Witten formulation.
The ghost number of $\PsiW$ is $1$ and the picture number of $\PsiW$ is $-1$.
Products $A \ast B$ of string fields $A$ and $B$ are defined by Witten's star product~\cite{Witten:bosonic},
$Q$ is the BRST operator, and $X_{\rm mid}$ is the picture-changing operator
inserted at the string midpoint.

As is mentioned in subsection~\ref{subsec:small/large}, 
the total picture number in a BPZ inner product in the small Hilbert space
has to be $-2$ for the inner product to be nonvanishing.
The picture-changing operator in the cubic interaction is inserted
to satisfy this condition.
However, the insertion of the picture-changing operator causes
the following two problems~\cite{Wendt}
originated from the singular OPE~(\ref{XX})
of the picture-changing operator with itself.

\begin{figure}
\begin{center}
\unitlength 0.1in
\begin{picture}( 26.2000,  9.1500)(  5.0000,-20.1600)
%
{\color[named]{Black}{%
\special{pn 13}%
\special{pa 932 1592}%
\special{pa 508 1166}%
\special{fp}%
\special{pa 1280 1450}%
\special{pa 996 1166}%
\special{fp}%
\special{pa 926 1592}%
\special{pa 500 2016}%
\special{fp}%
\special{pa 1280 1734}%
\special{pa 996 2016}%
\special{fp}%
\special{pa 1280 1450}%
\special{pa 2270 1450}%
\special{fp}%
\special{pa 1280 1734}%
\special{pa 2270 1734}%
\special{fp}%
}}%
%
{\color[named]{Black}{%
\special{pn 13}%
\special{pa 2688 1592}%
\special{pa 3114 1166}%
\special{fp}%
\special{pa 2342 1450}%
\special{pa 2624 1166}%
\special{fp}%
\special{pa 2696 1592}%
\special{pa 3120 2016}%
\special{fp}%
\special{pa 2342 1734}%
\special{pa 2624 2016}%
\special{fp}%
\special{pa 2342 1450}%
\special{pa 1350 1450}%
\special{fp}%
\special{pa 2342 1734}%
\special{pa 1350 1734}%
\special{fp}%
}}%
%
{\color[named]{Black}{%
\special{pn 4}%
\special{sh 1}%
\special{ar 1138 1592 10 10 0  6.28318530717959E+0000}%
\special{sh 1}%
\special{ar 2490 1592 10 10 0  6.28318530717959E+0000}%
}}%
\put(23.4100,-15.9100){\makebox(0,0){$X$}}%
\put(12.8600,-15.9100){\makebox(0,0){$X$}}%
%
{\color[named]{Black}{%
\special{pn 13}%
\special{pa 1138 1592}%
\special{pa 1280 1450}%
\special{dt 0.045}%
\special{pa 1280 1734}%
\special{pa 1138 1592}%
\special{dt 0.045}%
\special{pa 1138 1592}%
\special{pa 926 1592}%
\special{dt 0.045}%
\special{pa 2342 1734}%
\special{pa 2342 1734}%
\special{dt 0.045}%
}}%
%
{\color[named]{Black}{%
\special{pn 13}%
\special{pa 2342 1734}%
\special{pa 2342 1734}%
\special{dt 0.045}%
\special{pa 2484 1592}%
\special{pa 2342 1734}%
\special{dt 0.045}%
\special{pa 2342 1450}%
\special{pa 2484 1592}%
\special{dt 0.045}%
\special{pa 2484 1592}%
\special{pa 2688 1592}%
\special{dt 0.045}%
}}%
%
{\color[named]{Black}{%
\special{pn 8}%
\special{pa 1846 1182}%
\special{pa 2484 1182}%
\special{fp}%
\special{sh 1}%
\special{pa 2484 1182}%
\special{pa 2416 1162}%
\special{pa 2430 1182}%
\special{pa 2416 1202}%
\special{pa 2484 1182}%
\special{fp}%
}}%
%
{\color[named]{Black}{%
\special{pn 8}%
\special{pa 1670 1180}%
\special{pa 1134 1180}%
\special{fp}%
\special{sh 1}%
\special{pa 1134 1180}%
\special{pa 1200 1200}%
\special{pa 1186 1180}%
\special{pa 1200 1160}%
\special{pa 1134 1180}%
\special{fp}%
}}%
\put(17.7500,-11.6600){\makebox(0,0){$s$}}%
%
{\color[named]{Black}{%
\special{pn 4}%
\special{sh 1}%
\special{ar 1130 1590 16 16 0  6.28318530717959E+0000}%
}}%
%
{\color[named]{Black}{%
\special{pn 4}%
\special{sh 1}%
\special{ar 2480 1590 16 16 0  6.28318530717959E+0000}%
}}%
\end{picture}%
\caption{A tree-level diagram for the four-point scattering.}
\label{4-point scattering}
\end{center}
\end{figure}
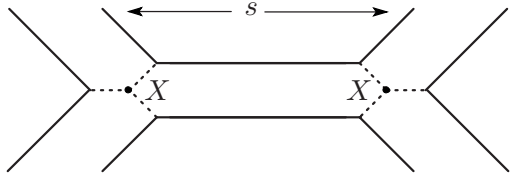

First, there are divergences in Feynman diagrams for scattering amplitudes.
Consider, for example, four-point amplitudes at the tree level.
We illustrated a Feynman diagram in figure~\ref{4-point scattering},
where $s$ is the Schwinger parameter for the propagator.
A picture-changing operator is inserted at each interaction point,
and two picture-changing operators collide
in the limit $s \to 0$.
Thus the amplitude diverges.
Note that this divergence exists at the tree level.

Second, the gauge variation of the action suffers from singularity.
The gauge transformation in the Witten formulation is given by
\begin{equation} \label{singular gauge transf}
\delta \PsiW = Q \Lambda^\mrm{W} 
+ g X_{\rm mid} \Bigl( \PsiW \ast \Lambda^\mrm{W} -\Lambda^\mrm{W} \ast \PsiW \Bigr) \,,
\end{equation}
where $\Lambda^\mrm{W}$ is a Grassmann-even gauge parameter
carrying the ghost number $0$ and the picture number $-1$.
The gauge variation of the action vanishes at $O(g)$.
At $O(g^2)$, however, two picture-changing operators collide,
and the gauge variation of the action is singular.

\subsection{The Berkovits formulation}
\label{Berkovits formulation}
The Berkovits formulation of open superstring field theory~\cite{Berkovits:NS}
is based on the large Hilbert space,
and no picture-changing operators are used.
The action $\SB$ in the NS sector takes the following
WZW-like form:\footnote{
A factor of the imaginary unit
for each term
is necessary for the action to be real.
See appendix B of~\cite{Torii:validity}.
}
\begin{equation} \label{S}
\SB = \frac{\iu}{2 g^2} \, \Bigl\langle  
\Ginv \bigl( QG\bigr) \,,\, \Ginv \bigl( \ez G\bigr)
\Bigr\rangle
-\frac{\iu}{2 g^2} \int^1_0 \diff t \,
\Bigl\langle
\bigl( \hatGinv \p_t \hatG \bigr) \,,\,
\Bigl\{ \hatGinv\bigl( Q\hatG\bigr), \hatGinv \bigl(\ez \hatG\bigr)\Bigr\}
\Bigr\rangle
\end{equation}
with 
\begin{equation} \label{G}
G = \exp \bigl( g\hs\Phi\bigr)\,,\quad
\hatG = \exp\bigl( t\hs g\hs\Phi\bigr)\,.
\end{equation}
Here $\Phi$ is the open superstring field in the NS sector, which is Grassmann even.
The ghost number of $\Phi$ is $0$ and the picture number of $\Phi$ is also $0$.
We often omit the symbol for the star product and write $A \ast B$ as $AB$ for simplicity,
but products of string fields are always
defined by Witten's star product.
The operators $Q$ and $\ez$ 
act as
derivations with respect to the star product, satisfying
\begin{equation} \label{Qeta} 
Q^2 = \ez^2 = \{ Q,\ez\} = 0\,.
\end{equation}

The action \eqref{S} is invariant under the gauge transformation parameterized by the Grassmann-odd string fields
$\LQ$ and $\Leta$ in the following form:
\begin{equation} \label{gauge transf}
\delta G = g\Bigl[ \bigl( Q\hs\LQ \bigr) G + G\bigl( \ez\hs\Leta\bigr)\Bigr]\,.
\end{equation}
We list the ghost number and the picture number
of string fields and gauge parameters in table~\ref{table:fields}.
\begin{table}
\begin{center}
\caption{The ghost number $\boldsymbol{g}$ and the picture number $\boldsymbol{p}$ of string fields and gauge parameters.}
\ \\[-.5ex]
{\renewcommand\arraystretch{1.3}
\begin{tabular}{|c||c|c|c||c|c|}
\hline
field & $\Phi$ & $\LQ$ & $\Leta$ & $\PsiW$ & $\Lambda^\mrm{W}$ \\
\hline
$(\boldsymbol{g}, \boldsymbol{p})$ & $(0,0)$ & $(-1,0)$ & $(-1,1)$ & $(1,-1)$ & $(0,-1)$ \\
\hline
\end{tabular}
}
\label{table:fields}
\end{center}
\end{table}
Compared to the Witten formulation,
the string field in the Berkovits formulation is in a larger space
and, to compensate it, the gauge symmetry is also larger
and parameterized by two string fields $\LQ$ and $\Leta$.

It has been confirmed at the tree level that
the correct four-point scattering amplitude
is reproduced in the Berkovits formulation~\cite{Berkovits:1999bs}. 
In section~\ref{section:regularization}, we will discuss how the problem of the divergence
in the four-point amplitude in the Witten formulation
is resolved in the Berkovits formulation.
We expand the action in the coupling constant $g$ as
\begin{equation}
\SB = \SB_2 + g \SB_3 + g^2 \SB_4 + O(g^3) \,,
\label{expansion:S} \\
\end{equation}
where
\bs
\begin{align}
\label{S0} 
\SB_2 &= -\frac{\iu}{2} \, \Bigl\langle \ez\Phi , Q \Phi \Bigr\rangle\,,\\[.5ex]
\label{S1} 
\SB_3 &= \frac{\iu}{3!} \, \Bigl\langle \ez\Phi , \bigl[\Phi, Q\Phi\bigr] \Bigr\rangle
\,,\\[.5ex]
\label{S2} 
\SB_4
&= -\frac{\iu}{4!} \, \Bigl\langle \ez\Phi , \Bigl[ \Phi, \bigl[\Phi, Q\Phi\bigr] \Bigr] \Bigr\rangle
= \frac{\iu}{24} \, \Bigl\langle \left[\Phi,\eta_0\Phi\right] , \left[\Phi,Q\Phi\right] \Bigr\rangle \,.
\end{align}
\es
The expansion of the gauge transformation in $g$ is
\begin{equation}
\delta \Phi 
= \left( Q \LQ + \ez \Leta \right)
- \frac{g}{2} \, \bigl[ \, \Phi, Q \LQ - \ez \Leta \, \bigr]
+ \frac{g^2}{12} \, \Bigl[ \, \Phi,\bigl[ \, \Phi, Q \LQ + \ez \Leta \, \bigr] \, \Bigr] +O(g^3) \,.
\label{expansion:delta}
\end{equation}
In the Witten formulation,
there are no terms whose orders are higher than $O(g)$
both in the action and in the gauge transformation.
As we will see in the following sections, higher-order terms in the Berkovits formulation
play an important role in solving the two problems in the Witten formulation
we mentioned in the preceding subsection.

\section{Partial gauge fixing}
\label{section:Gauge-fixing}
\setcounter{equation}{0}
The Witten formulation and the Berkovits formulation of open superstring field theory
may look rather different.
In the free theory, however, we can demonstrate that the two formulations are equivalent
and are related via partial gauge fixing:\footnote{
For the correspondence between the two formulations in the free theory
via complete gauge fixing, 
see~\cite{Kroyter:2012ni}.
}
the Witten formulation can be obtained from the Berkovits formulation
by fixing the gauge degrees of freedom associated with the gauge parameter 
$\Leta$ in~\eqref{gauge transf}.
In the interacting theory, we introduce a one-parameter family of gauge choices
for the partial gauge fixing, and we demonstrate that the Witten formulation
can be obtained from the Berkovits formulation in a singular limit
of the one-parameter family of gauges.
We can then use the Berkovits formulation as a regularization of the Witten formulation
and discuss the nature of the singularity in the Witten formulation.

The idea of using partial gauge fixing
to relate the Witten formulation and the Berkovits formulation is not new.
See~\cite{review:Berkovits, constrainedBV}, for example.
The main point of this paper is our explicit construction
of the one-parameter family of gauges, which enables us
to discuss the singularity of the Witten formulation
in a well-defined setting.
Partial gauge fixing can be subtle in quantum theory,
where ghosts associated with gauge fixing are required.
We restrict ourselves to the situations
where no such subtleties appear.
In section~\ref{section:regularization}, we discuss how the singularity
in four-point amplitudes in the Witten formulation
can be resolved using the Berkovits formulation
as a regularization.
We only consider tree-level amplitudes, where no ghosts propagate,
and we anyway have to fix gauge completely for calculations of scattering amplitudes.
In section~\ref{sec:relation}, we discuss how the singularity
in the gauge variation of the action in the Witten formulation
to the second order in the open string coupling constant can be resolved
using the idea of partial gauge fixing.
We do not see any subtleties associated with partial gauge fixing
in this discussion.

\subsection{The strategy}
\label{subsec:idea}
Let us begin with the free theory and demonstrate the equivalence
of the Witten formulation and the Berkovits formulation.
The equation of motion in the Witten formulation is
\begin{equation} \label{freeEOM:Witten}
Q\PsiW =0 \,,
\end{equation}
where the string field $\PsiW$ is in the small Hilbert space:
\begin{equation}
\eta_0 \PsiW = 0 \,.
\end{equation}
The equation of motion in the Berkovits formulation is
\begin{equation} \label{freeEOM:Berkovits}
Q\ez\Phi = 0\,.
\end{equation}
The gauge transformation~(\ref{gauge transf}) reduces to
\begin{equation}
\delta \Phi = Q \LQ + \ez \Leta
\label{linearized-gauge-transformation}
\end{equation}
in the free theory.
Using the relation
\begin{equation}
\{\xz, \ez\} =1\,,
\end{equation}
we can rewrite \eqref{freeEOM:Witten} as
\begin{equation}
0 = Q\PsiW = Q\{\xz, \ez\}\PsiW = Q\ez\bigl(\xz\PsiW\bigr) \,. 
\end{equation}
Therefore, for any solution $\PsiW$ in the Witten formulation,
we have a solution given by $\xz\PsiW$ in the Berkovits formulation.

On the other hand, any string field $\Phi$ in the Berkovits formulation
can be brought to the form
\begin{equation} \label{XiPsi}
\Phi = \xz\Psi \quad \text{with} \quad \eta_0 \Psi = 0
\end{equation}
by the gauge transformation with the parameter $\Leta$
in~\eqref{linearized-gauge-transformation} as
\begin{equation}
\Phi \to \Phi + \ez \Leta \quad \text{with} \quad \Leta = -\xz \Phi
\end{equation}
because
\begin{equation}
\Phi = \{\xz, \ez\} \Phi = \xz \Psi -\ez \Leta
\end{equation}
under the identification
\begin{equation}
\Psi = \ez \Phi \,.
\end{equation}
The condition for this partial gauge fixing can be stated as
\begin{equation} \label{XiPhi}
\xz\Phi = 0\,,
\end{equation}
and $\Phi$ satisfying this condition can be written
in the form~(\ref{XiPsi}).

Let us next evaluate the action
\begin{equation}
\SB_2 = -\frac{\iu}{2} \, \bigl\langle \ez\Phi, Q \Phi \bigr\rangle
\end{equation}
in the Berkovits formulation under the partial gauge fixing.
When $\Phi$ is written in the form~(\ref{XiPsi}), we have
\begin{equation}
\ez \Phi = \Psi \,, \qquad
Q \Phi = Q \xz \Psi = -\xz Q \Psi + X_0 \Psi \,,
\end{equation}
where $X_0$ is the zero mode of the picture-changing operator:
\begin{equation}
X_0 = \{ Q, \xz \} \,.
\end{equation}
Note that $X_0$ commutes with $\ez$ because
\begin{equation}
[ \, \ez, X_0 \, ] = [ \, \ez, \{ Q, \xz \} \, ]
= {}-[ \, Q, \{ \xz, \ez \} \, ] -[ \, \xz, \{ \ez, Q \} \, ] = 0 \,.
\end{equation}
Therefore, the string field $X_0 \Psi$ is annihilated by $\ez$
and in the small Hilbert space.
It follows from~\eqref{basic_correlator} that
the BPZ inner product
$\langle \, A, B \, \rangle$
vanishes when $A$ and $B$ are both in the small Hilbert space.
In the BPZ inner product of
$\langle \, \ez \Phi, Q \Phi \, \rangle$,
the term $X_0 \Psi$ in $Q \Phi$ does not contribute, and we have
\begin{equation}
-\frac{\iu}{2} \, \bigl\langle \, \ez\Phi, Q \Phi \, \bigr\rangle
= \frac{\iu}{2} \, \bigl\langle \, \Psi, \xz Q \Psi \, \bigr\rangle
= -\frac{1}{2} \, \bllangle \, \Psi, Q\Psi \, \brrangle\,.
\end{equation}
This coincides with the action in the Witten formulation under the identification
\begin{equation} \label{identification}
\Psi \cong \PsiW\,.
\end{equation}
We have thus seen that the action in the Berkovits formulation
reduces to the gauge-invariant action in the Witten formulation
under the partial gauge fixing.

So far we have only used the properties of $\xz$
that $\xz^2 = 0$ and $\{ \xz, \ez \} = 1$.
We can therefore replace $\xz$ with any Grassmann-odd operator $\Xi$
satisfying
\begin{equation}
\Xi^2 = 0\,, \quad \{\Xi, \ez\} = 1\,,
\end{equation}
and we consider the condition for partial gauge fixing given by\footnote{
The compatibility of this condition
with the reality condition on the string field \cite{Gaberdiel:1997ia}
imposes a constraint on $\Xi$.
We will discuss this constraint in appendix~\ref{section:reality}.} 
\begin{equation} \label{Xi_gauge_condition} 
\Xi \Phi = 0 \,.
\end{equation}
We consider $\Xi$ carrying the same quantum numbers as
$\xi_0$:
the ghost number of $\Xi$ is $-1$ and the picture number of $\Xi$ is $1$.
The string field satisfying this condition can be written as
\begin{equation}
\Phi = \Xi \Psi \quad \text{with} \quad \ez \Psi = 0 \,.
\end{equation}
The relation~\eqref{small/large:BPZ} generalizes to 
\begin{equation} \label{small/large:BPZ:Xi}
\llangle \, A,B \, \rrangle = \iu\hs\langle \, \Xi A,\hs B \, \rangle \,,\quad
\llangle \, A,B \, \rrangle = \iu\hs (-1)^A \hs\langle \, A ,\hs \Xi B \, \rangle
\end{equation}
for any $A$ and $B$ in the small Hilbert space,
and the action in the Berkovits formulation under the partial gauge fixing
reduces to the action in the Witten formulation
with the identification $\Psi \cong \PsiW$.

Let us now consider the interacting theory.\footnote{
We can impose the condition \eqref{Xi_gauge_condition} on $\Phi$
in the interacting theory as well,
using the gauge degrees of freedom associated with $\Leta$.
See~\cite{IT} for details.}
If we choose $\Xi$ to be
\begin{equation} \label{xi:midpoint}
\Xi = \xi_{\rm mid}\,, 
\end{equation}
where $\xi_{\rm mid}$ is the midpoint insertion of $\xi$,
we naively expect that the cubic interaction~\eqref{S1} in the Berkovits formulation
will reduce
to that in the Witten formulation 
for
the following reason.
The cubic interaction~\eqref{S1} consists of the three string fields
$\Phi$, $\ez \Phi$, and $Q \Phi$.
When $\Phi = \xi_{\rm mid} \Psi$, these three string fields are
\begin{equation}
\Phi = \xi_{\rm mid} \Psi \,, \qquad
\ez \Phi = \Psi \,, \qquad
Q \Phi = -\xi_{\rm mid} Q \Psi + X_{\rm mid} \Psi \,.
\end{equation}
In the cubic interaction 
defined by
the star product,
the midpoints of the three string fields are mapped to the same point.
Thus the term $-\xi_{\rm mid} Q \Psi$ in $Q \Phi$ does not contribute
because the insertion of $\xi$ from this term
and the insertion of $\xi$ from $\Phi$
are mapped to the same point
and $\xi (z)^2 = 0$.
The cubic interaction then consists of $\xi_{\rm mid} \Psi$,
$\Psi$, and $X_{\rm mid} \Psi$,
and the operator $\xi_{\rm mid}$ 
will be used
to saturate the zero mode of $\xi$.
Written in terms of the BPZ inner product in the small Hilbert space,
the cubic interaction formally reduces to the cubic term
in the Witten formulation~(\ref{Witten:NS-action}).
Actually, this argument is formal
because the insertion of $X$ and the insertion of $\xi$ are mapped to the same point 
but the OPE of $X$ and $\xi$ is singular:
\begin{equation}
X(z_1) \, \xi(z_2)
\sim-\frac{2}{(z_1-z_2)^2} \, b\e^{2\phi}(z_2)
-\frac{1}{z_1-z_2} \, \partial\big(b\e^{2\phi}\big)(z_2) \,.
\label{X-xi-OPE}
\end{equation}
If we regularize the condition $\xi_{\rm mid} \Phi = 0$, however,
we expect to obtain the cubic interaction of the Witten formulation
in the limit where the condition $\xi_{\rm mid} \Phi = 0$ is recovered.
We will explicitly demonstrate this 
in subsection~\ref{section:vertices}
after providing a one-parameter family of gauge conditions
in the next subsection.

Let us also consider the quartic interaction \eqref{S2} in the Berkovits formulation.
It consists of two string fields of $\Phi$, $\ez \Phi$, and $Q \Phi$.
When $\Phi = \xi_{\rm mid} \Psi$,
the term $-\xi_{\rm mid} Q \Psi$ in $Q \Phi$ again does not contribute
because this time three insertions of $\xi$
are mapped to the same point.
The quartic interaction then consists of
two string fields of $\xi_{\rm mid} \Psi$,
$\Psi$, and $X_{\rm mid} \Psi$,
and two insertions of $\xi$ and one insertion of $X$
are mapped to the same point.
Assuming an appropriate regularization,
we use one of the two insertions of $\xi$ to saturate the zero mode of $\xi$,
but we still expect a singularity from the collision
of the remaining $\xi$ and $X$.
We thus anticipate that the quartic interaction in the Berkovits formulation
diverges in the limit where the cubic interaction reduces
to that in the Witten formulation.
Using the one-parameter family of gauges we introduce in the next subsection, 
we will discuss the nature of this singular limit
and see the role of the divergent quartic interaction.

\subsection{A one-parameter family of conditions for partial gauge fixing}
\label{subsec:gfc}
We have seen that the choice $\Xi = \xi_{\rm mid}$ is singular.
How should we regularize it?
Using the state-operator correspondence in conformal field theory (CFT),
an open string state can be described as a state defined on the unit semi-circle
in the upper half-plane of $z$
by the path integral in the interior region $| z | < 1$
with the corresponding operator inserted at the origin.
In this description, the open string midpoint corresponds to the point $z = \iu$,
and $\xi_{\rm mid}$ is given by
\begin{equation}
\begin{split}
\xi_{\rm mid} = \xi (\iu) & = \sum_{n = -\infty}^\infty (-\iu \, )^n \, \xi_n \\
& =\xi_0
-\iu \, ( \xi_1 -\xi_{-1} ) -( \xi_2 +\xi_{-2} )
+\iu \, ( \xi_3 -\xi_{-3} ) +( \xi_4 +\xi_{-4} ) + \ldots \,.
\end{split}
\end{equation}
How about regularizing $\xi_{\rm mid}$
by slightly changing the location of the operator from $z = \iu$
to $z = \iu \e^{-\lambda}$
with $\lambda$ being real and positive?
The resulting operator $\xi ( \iu \e^{-\lambda} )$ is expanded as follows:
\begin{equation}
\begin{split}
\xi ( \, \iu \e^{-\lambda} )
& = \sum_{n = -\infty}^\infty \frac{\xi_n}{( \, \iu \e^{-\lambda} )^n} \\
& =\xi_0
-\iu \, ( \e^\lambda \, \xi_1 -\e^{-\lambda} \, \xi_{-1} )
-( \e^{2 \lambda} \, \xi_2 +\e^{-2 \lambda} \, \xi_{-2} )
+\iu \, ( \e^{3 \lambda} \, \xi_3 -\e^{-3 \lambda} \, \xi_{-3} )
+ \ldots \,.
\end{split}
\label{xi-off-midpoint}
\end{equation}
The singularity when we use $\xi_{\rm mid}$ for the partial gauge fixing is regularized
and $\xi_{\rm mid}$ is recovered in the limit $\lambda \to 0$.
However, the problem with this choice is that
the location of the operator is in the region of the path integral.

One way to solve this problem is to consider an operator
defined by an integral along the unit circle $C$
instead of an insertion of a local operator.\footnote{
A similar idea was used in~\cite{Drukker:2005hr} to regularize the midpoint $c$-ghost insertion 
in the kinetic term of vacuum string field theory~\cite{Rastelli:2000hv, Rastelli:2001jb, Rastelli:2001uv}. 
}
We define $\Xil$ by
\begin{equation} \label{Xi,u}
\Xil = \oint_C \frac{\diff z}{2\pi\iu} \, u_\lambda (z) \, \xi(z)
\end{equation}
with
\begin{equation} \label{u:even}
u_\lambda (z) = \frac{1}{z-\iu\e^{-\lambda}} - \frac{1}{z-\iu\e^{\lambda}} \,,
\end{equation} 
where the parameter $\lambda$ is in the region $0 < \lambda < \infty$.
If we deform the contour $C$ to a contour $C'$
which encircles the origin counterclockwise 
along a circle with its radius smaller than $\e^{-\lambda}$,
the contribution from the pole of $u_\lambda (z)$ at $z = \iu \e^{-\lambda}$
gives $\xi ( \, \iu \e^{-\lambda} )$
and the operator $\Xil$ is expressed as
\begin{equation}
\Xil = \xi ( \, \iu \e^{-\lambda} )
+\oint_{C'} \frac{\diff z}{2\pi\iu} \, u_\lambda (z) \, \xi(z) \,. 
\end{equation}
In the region $|z| < \e^{-\lambda}$, 
$u_\lambda (z)$ does not have any singularity
and it vanishes in the limit $\lambda \to 0$.
We thus find
\begin{equation}
\Xil = \xi ( \, \iu \e^{-\lambda} ) +O (\lambda) \,.
\label{Xil insertion}
\end{equation}
Note that the contribution from the second term of $u_\lambda (z)$
is necessary to make the integral along $C'$ vanish in the limit $\lambda \to 0$.
We can also confirm~(\ref{Xil insertion}) more directly.
In the annulus region $\e^{-\lambda} < | z | < \e^\lambda$,
the Laurent expansion of $u_\lambda (z)$ in $z$ is given by
\begin{equation}
u_\lambda (z) = \frac{1}{z}
+\sum_{k=1}^\infty \, \frac{( \, \iu \e^{-\lambda} )^k}{z^{k+1}}
+\sum_{k=1}^\infty \, ( -\iu \e^{-\lambda} )^k \, z^{k-1} \,.
\end{equation}
The mode expansion of $\Xil$ is then
\begin{equation}
\begin{split}
\Xil & = \xi_0
+\sum_{k=1}^\infty \, ( \, \iu \e^{-\lambda} )^k \, \xi_{-k}
+\sum_{k=1}^\infty \, ( -\iu \e^{-\lambda} )^k \, \xi_k \\
& =\xi_0
-\iu \e^{-\lambda} \, ( \xi_1 -\xi_{-1} )
-\e^{-2 \lambda} \, ( \xi_2 +\xi_{-2} )
+\iu \e^{-3 \lambda} \, ( \xi_3 -\xi_{-3} )
+ \ldots \,. \\
\end{split}
\label{Xil-expansion} 
\end{equation}
Comparing this expansion with~(\ref{xi-off-midpoint}),
we verify~(\ref{Xil insertion}).

To summarize, we have explicitly constructed $\Xil$ in~(\ref{Xi,u})
labeled by the parameter $\lambda$
in the range $0 < \lambda < \infty$
such that it reduces to $\xi_{\rm mid}$ in the limit $\lambda \to 0$.
We use this operator to impose the condition
\begin{equation}
\Xil \Phi = 0
\label{gauge-condition}
\end{equation}
for the partial gauge fixing,
and the string field satisfying this condition
can be written as
\begin{equation}
\Phi = \Xil \Psi \quad \text{with} \quad \ez \Psi = 0 \,.
\end{equation}
The operator $\Xil$ is BPZ even, which can be easily seen from the expansion~(\ref{Xil-expansion})
because the BPZ conjugate of $\xi_n$ is $(-1)^n \, \xi_{-n}$,
and we show in appendix~\ref{section:reality}
that the condition~(\ref{gauge-condition}) is compatible
with the reality condition on $\Phi$.
In the opposite limit $\lambda \to \infty$,
$\Xil$ reduces to $\xi_0$.
The operator $\Xil$ therefore interpolates between $\xi_{\rm mid}$ and $\xi_0$.

Another important operator is the BRST transformation of $\Xil$.
We denote it by ${\cal X}_\lambda$:
\begin{equation} \label{calX}
\calX_\lambda := \{Q, \Xi_\lambda\}\,.
\end{equation}
It follows from~(\ref{Xil insertion}) that
\begin{equation} \label{X_lambda insertion}
\calX_\lambda = X(\iu\e^{-\lambda}) +O(\lambda) \,.
\end{equation}
The operator approaches the midpoint insertion $X_{\rm mid}$
of the local picture-changing operator in the limit $\lambda \to 0$
and it reduces to $X_0$ in the limit $\lambda \to \infty$.

\subsection{Interaction terms under the partial gauge fixing}
\label{section:vertices}
Now that we have the condition~\eqref{gauge-condition}
for the partial gauge fixing
labeled by $\lambda$ in the range $0 < \lambda < \infty$,
we can discuss the singular limit $\lambda \to 0$ in a well-defined setting.
Ingredients of the interaction terms in the Berkovits formulation are $\Phi$, $\ez \Phi$, and $Q \Phi$.
Under the partial gauge fixing, the string field $\Phi$ can be written as
\begin{equation}
\Phi = \Xil \Psi \,,
\end{equation}
where $\Psi$ is in the small Hilbert space.
The string fields $\ez \Phi$ and $Q \Phi$ are
\begin{equation}
\ez \Phi = \Psi \,, \qquad
Q \Phi = Q \Xil \Psi \,.
\end{equation}
When we discuss the limit $\lambda \to 0$,
it is useful to write $Q \Phi$ as follows:
\begin{equation}
Q \Phi = Q \Xil \Psi = -\Xil Q \Psi +\Xl \Psi \,.
\end{equation}
Note that $\Xl \Psi$ is in the small Hilbert space
because $\Xl$ commutes with $\ez$.
When we denote the operator corresponding to the state $\Psi$ by $\Psi (0)$,
the operator $\Xil \cdot \Psi (0)$ corresponding to the state $\Xil \Psi$
can be written as
\begin{equation}
\Xil \cdot \Psi (0) = \xi ( \iu\e^{-\lambda} ) \, \Psi (0) + O(\lambda) \,,
\label{Xil-Psi-approximation}
\end{equation}
where we used~(\ref{Xil insertion}).
Similarly, the operator corresponding to $\Xil Q \Psi$ can be written as
\begin{equation}
\xi ( \iu\e^{-\lambda} ) \, Q \cdot \Psi (0) + O(\lambda) \,,
\label{Xil-Q-Psi-approximation}
\end{equation}
and the operator corresponding to $\Xl \Psi$ can be written as
\begin{equation}
X ( \iu\e^{-\lambda} ) \, \Psi (0) + O(\lambda) \,,
\label{Xl-Psi-approximation}
\end{equation}
where we used~(\ref{X_lambda insertion}).

\begin{figure}
\begin{center}
\includegraphics[width=100mm]{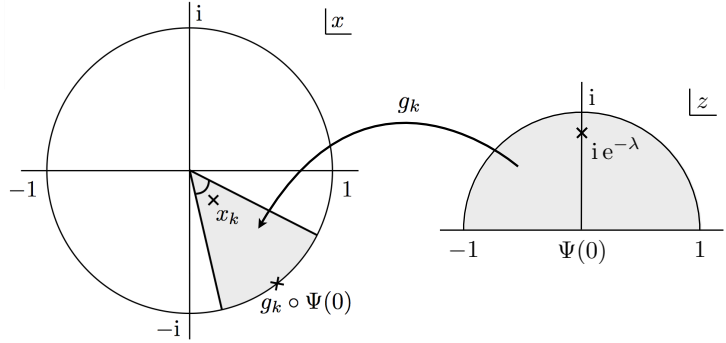}
\end{center}
\caption{The conformal transformation $g_k (z)$ maps
the upper half-disk to the wedge region shown in the figure.
When $\lambda$ is small, the point $z = \iu\e^{-\lambda}$
near the open-string midpoint in the upper half-disk
is mapped to the point $\dc_k$ near the origin.}
\label{fig:inot1_vertex}
\end{figure}

The $n$-point interactions in the Berkovits formulation
are constructed from these ingredients
and consist of terms of the form
\begin{equation} \label{term_in_n-point}
\Big\langle\big(Q\Phi\big), \Phi^m\big(\ez\Phi\big)\Phi^{n-m-2}\Big\rangle
\end{equation}
with $0 \leq m \leq n-2$.
Under the partial gauge fixing, this can be written as
\begin{equation}
\Big\langle\bigl(Q\Xil\Psi\bigr), \bigl(\Xil\Psi\bigr)^m\Psi\bigl(\Xil\Psi\bigr)^{n-m-2}\Big\rangle \,.
\end{equation}
To discuss the limit $\lambda \to 0$, we write
\begin{align} \label{n-pt:xi}
&
\Big\langle\bigl(Q\Xil\Psi\bigr), \bigl(\Xil\Psi\bigr)^m\Psi\bigl(\Xil\Psi\bigr)^{n-m-2}\Big\rangle \nonumber \\*
=\ &
\Big\langle\big(\Xl\Psi\big), \big(\Xil\Psi\big)^m\Psi\big(\Xil\Psi\big)^{n-m-2}\Big\rangle
-\Big\langle\big(\Xil Q\Psi\big), \big(\Xil\Psi\big)^m\Psi\big(\Xil\Psi\big)^{n-m-2}\Big\rangle\,.
\end{align}
It is convenient to use CFT correlation functions to express these terms.
An expression using a CFT correlation function on the unit disk $D_2$
for $\langle \, A_1, A_2 \, A_3 \, \ldots \, A_n \, \rangle$ is
\begin{equation}
\langle \, A_1, A_2 \, A_3 \, \ldots \, A_n \, \rangle
= \bigl\langle \, \bigl( g_1 \circ\ A_1 (0) \bigr) \bigl( g_2 \circ A_2 (0) \bigr)
\bigl( g_3 \circ A_3 (0) \bigr) \ldots \bigl( g_n \circ A_n (0)\bigr) \, \bigr\rangle_{D_2} \,,
\end{equation}
where $A_k (0)$ is the operator corresponding to the state $A_k$
and $g_k \circ A_k (0)$ is the conformal transformation of $A_k (0)$
under the map $g_k (z)$ given by
\begin{equation} \label{mapping}
z \mapsto g_k(z) = \e^{\frac{2\pi\iu}{n}k}
\left(\frac{1+\iu\lc}{1-\iu\lc}\right)^{\frac{2}{n}} \quad \text{for} \quad k=1, \ldots , n \,.
\end{equation}
See figure~\ref{fig:inot1_vertex}.
In our convention, the conformal transformation $f \circ \varphi (z)$ is given by
$f \circ \varphi (z) = ( f' (z) )^h \, \varphi ( f(z) )$
when $\varphi (z)$ is a primary field of conformal weight $h$.
If we neglect terms of $O(\lambda)$ in~(\ref{Xil-Psi-approximation}),
(\ref{Xil-Q-Psi-approximation}), and~(\ref{Xl-Psi-approximation}),
the first term on the right-hand side of \eqref{n-pt:xi}
up to an overall sign factor is given by
\begin{equation} \label{n-point_first_correlation}
\Big\langle X(\dc_1)\xi(\dc_2)\xi(\dc_3)\ldots\widehat{\xi(\dc_{m+2})}\ldots\xi(\dc_n) 
\prod_{k=1}^ng_k\circ\Psi(0)\Big\rangle_{D_2}\,,
\end{equation}
where
\begin{equation}
\dc_k = g_k (\iu\e^{-\lambda})
\label{dc_k} 
\end{equation}
and
\begin{equation}
\prod_{k=1}^n g_k\circ\Psi(0)
= \Bigl( g_1\circ\Psi(0)\Bigr)\Bigl( g_2\circ\Psi(0)\Bigr)
\ldots \Bigl( g_n\circ\Psi(0)\Bigr) \,.
\end{equation}
The hat on $\xi(\dc_{m+2})$ indicates that $\xi(\dc_{m+2})$ is omitted.
Note that $X (z)$ and $\xi (z)$ are primary fields
of conformal weight $0$
so there are no associated conformal factors. 
Similarly, the second term on the right-hand side of~\eqref{n-pt:xi}
in the same approximation is given by
\begin{equation} \label{n-point_second_correlation}
\Big\langle \xi(\dc_1)\xi(\dc_2) \ldots\widehat{\xi(\dc_{m+2})}\ldots\xi(\dc_n) \,
g_1 \circ \bigl( Q \cdot \Psi (0) \bigr) \prod_{k=2}^n g_k\circ\Psi(0)\Big\rangle_{D_2}
\end{equation}
up to an overall sign factor.
In the limit $\lambda \to 0$, the point $\dc_k$ approaches the origin:
\begin{equation}
\dc_k \to 0 \quad \text{as} \quad \lambda \to 0 \,.
\end{equation}
It follows that the correlation function~(\ref{n-point_second_correlation})
vanishes in this limit because multiple insertions of $\xi$ collide.
Therefore, we can safely neglect
the second term on the right-hand side of~\eqref{n-pt:xi}
in the limit $\lambda \to 0$.
On the other hand, the correlation function~(\ref{n-point_first_correlation})
contains an insertion of the picture-changing operator $X$.
Since the OPE of $X$ with $\xi$ is singular, we need detailed analysis
in the limit $\lambda \to 0$.

To summarize, the $n$-point interactions in the Berkovits formulation
under the partial gauge fixing 
are given approximately by terms expressed as~(\ref{n-point_first_correlation})
when $\lambda$ is small.
In other words, these terms consist of
$\langle \, \Psi, \Psi^{n-1} \, \rangle$
with one insertion of the picture-changing operator $X$
and $n-2$ insertions of $\xi$ near the open string midpoint.
In the limit $\lambda \to 0$, all these operator insertions
approach the midpoint and could cause divergences.
However, all the expressions are well defined and finite
as long as $\lambda$ is finite.

Let us now look at the cubic interaction $\SB_3$ in~(\ref{S1}).
Under the partial gauge fixing, this can be written as
\begin{equation}
\SB_3 =\frac{\iu}{3!} \biggl(\Big\langle\Psi, \big(\Xil\Psi\big)\big(Q\Xil\Psi\big)\Big\rangle
-\Big\langle\Psi, \big(Q\Xil\Psi\big)\big(\Xil\Psi\big)\Big\rangle\biggr) \,.
\label{cubic-regularized}
\end{equation}
When $\lambda$ is small, we have
\begin{align}
\SB_3 & =\frac{\iu}{3!} \biggl(\Big\langle\Psi, \big(\Xil\Psi\big)\big(\Xl\Psi\big)\Big\rangle
-\Big\langle\Psi, \big(\Xl\Psi\big)\big(\Xil\Psi\big)\Big\rangle \nonumber \\
& \qquad \quad~
-\Big\langle\Psi, \big(\Xil\Psi\big)\big(\Xil Q\Psi\big)\Big\rangle
+\Big\langle\Psi, \big(\Xil Q\Psi\big)\big(\Xil\Psi\big)\Big\rangle\biggr) 
\nonumber \\
& \simeq \frac{\iu}{3!} \biggl(\Big\langle\Psi, \big(\Xil\Psi\big)\big(\Xl\Psi\big)\Big\rangle
-\Big\langle\Psi, \big(\Xl\Psi\big)\big(\Xil\Psi\big)\Big\rangle \biggr) \,.
\label{S1_in_Xie_gauge}
\end{align}
Here and in what follows we use the notation $A \simeq B$
when $A = B$ up to terms which vanish in the limit $\lambda \to 0$. 
We dropped
the two terms in the second line of~(\ref{S1_in_Xie_gauge}), 
following the argument that the second term
on the right-hand side of~(\ref{n-pt:xi}) can be neglected in the limit $\lambda \to 0$.
Using CFT correlation functions on the unit disk $D_2$,
we can express these terms as
\begin{equation}
\SB_3 \simeq -\frac{\iu}{3!} \,
\Big\langle \, \xi (\dc_2) \, X (\dc_3) 
\prod_{k=1}^3 g_k\circ\Psi(0)\Big\rangle_{D_2}
-\frac{\iu}{3!} \,
\Big\langle X (\dc_2) \, \xi (\dc_3)
\prod_{k=1}^3 g_k\circ\Psi(0)\Big\rangle_{D_2}\,,
\label{S1_CFT}
\end{equation}
where we further dropped terms of $O(\lambda)$
in~(\ref{Xil-Psi-approximation}) and~(\ref{Xl-Psi-approximation}).
In each of these two terms, the insertion of $X$ and the insertion of $\xi$
approach the origin and collide in the limit $\lambda \to 0$.
While the OPE~(\ref{X-xi-OPE}) of $X$ and $\xi$ is singular,
the singular terms do not saturate the zero mode of $\xi$
and do not contribute in the correlation functions.
We can thus take the limit $\lambda \to 0$ to obtain
\begin{equation}
\SB_3 \Bigr|_{\lambda\to 0}
= -\frac{\iu}{3} \,
\Big\langle : \! \xi X \! : \! (0) 
\prod_{k=1}^3 g_k\circ\Psi(0)\Big\rangle_{D_2}
= -\frac{1}{3} \,
\Bllangle X(0) \prod_{k=1}^3 g_k\circ\Psi(0)\Brrangle_{D_2} \,,
\end{equation}
where we wrote the normal-ordering symbol explicitly.
This is precisely the cubic interaction in the Witten formulation.
We can also obtain this result in the following way.
The expression in the third line of~(\ref{S1_in_Xie_gauge})
can be transformed as
\begin{equation}
\begin{split}
\SB_3 & \simeq -\frac{\iu}{3!} \, \biggl(
\Big\langle \, \Xil \Psi, \, \big(\Xl\Psi\big) \Psi \, \Big\rangle
+\Big\langle \, \Xil \Psi, \, \Psi \big(\Xl\Psi\big) \, \Big\rangle \biggr) \\
& = -\frac{1}{3!} \, \biggl(
\Bllangle \, \Psi, \, \big(\Xl\Psi\big) \Psi \, \Brrangle
+\Bllangle \, \Psi, \, \Psi \big(\Xl\Psi\big) \, \Brrangle \biggr) \,,
\end{split}
\end{equation}
where we used~(\ref{small/large:BPZ:Xi}).
Since $\Xl \to X_{\rm mid}$ in the limit $\lambda \to 0$,
we obtain the cubic term in the Witten formulation:
\begin{equation}
\SB_3 \Bigr|_{\lambda\to 0}
= -\frac{1}{3} \, \Bllangle \, \Psi, \, X_{\rm mid} \big( \Psi^2 \big) \, \Brrangle \,,
\end{equation}
where we used the relations
\begin{equation}
\big( X_{\rm mid} A \, \big) \, B = A \, \big( X_{\rm mid} B \, \big)
= X_{\rm mid} \big( A B \, \big)
\label{X_mid-identities} 
\end{equation}
for any pair of string fields $A$ and $B$.
While $\SB_3$ in the limit $\lambda \to 0$ is finite,
we need to keep $\lambda$ finite when we discuss
the on-shell four-point amplitude and the gauge variation of the action,
and we will use expressions such as~(\ref{S1_in_Xie_gauge}) or~(\ref{S1_CFT})
as our regularization of the cubic term in the Witten formulation.

Let us next consider the quartic interaction~$\SB_4$ in~(\ref{S2}).
It turns out to be useful to transform $\SB_4$ in the following way:
\begin{equation}
\begin{split}
\SB_4 & = \frac{\iu}{24} \,
\Bigl\langle \Phi^2, \left(Q\Phi\right)\left(\eta_0\Phi\right) \Bigr\rangle
-\frac{\iu}{24} \,
\Bigl\langle \Phi^2, \left(\eta_0\Phi\right)\left(Q\Phi\right) \Bigr\rangle
-\frac{\iu}{12} \,
\Bigl\langle \Phi \left(Q\Phi\right), \Phi \left(\eta_0\Phi\right) \Bigr\rangle \\
& = \frac{\iu}{8} \,
\Bigl\langle \Phi^2, \left(Q\Phi\right)\left(\eta_0\Phi\right) \Bigr\rangle
-\frac{\iu}{8} \,
\Bigl\langle \Phi^2, \left(\eta_0\Phi\right)\left(Q\Phi\right) \Bigr\rangle
+\frac{\iu}{12} \,
\Bigl\langle \Phi^3,  Q\eta_0\Phi \Bigr\rangle \,.
\end{split}
\label{quartic-useful} 
\end{equation}
Under the partial gauge fixing, this can be written as
\begin{equation}
\SB_4 = \frac{\iu}{8} \,
\Bigl\langle \left(\Xil\Psi\right)^2, \left(Q\Xil\Psi\right)\Psi \Bigr\rangle
-\frac{\iu}{8} \,
\Bigl\langle \left(\Xil\Psi\right)^2, \Psi\left(Q\Xil\Psi\right) \Bigr\rangle
+\frac{\iu}{12} \,
\Bigl\langle \left(\Xil\Psi\right)^3,  Q\Psi \Bigr\rangle \,.
\label{quartic-regularized}
\end{equation}
Let us investigate the limit $\lambda \to 0$.
The last term on the right-hand side vanishes in this limit
because three insertions of $\xi$ collide. We thus have
\begin{equation}
\begin{split}
\SB_4 & = \frac{\iu}{8} \,
\Bigl\langle \left(\Xil\Psi\right)^2, \left(\Xl\Psi\right)\Psi \Bigr\rangle
-\frac{\iu}{8} \,
\Bigl\langle \left(\Xil\Psi\right)^2, \Psi\left(\Xl\Psi\right) \Bigr\rangle \\
& \quad ~
-\frac{\iu}{8} \,
\Bigl\langle \left(\Xil\Psi\right)^2, \left(\Xil Q\Psi\right)\Psi \Bigr\rangle
+\frac{\iu}{8} \,
\Bigl\langle \left(\Xil\Psi\right)^2, \Psi\left(\Xil Q\Psi\right) \Bigr\rangle
+\frac{\iu}{12} \,
\Bigl\langle \left(\Xil\Psi\right)^3,  Q\Psi \Bigr\rangle \\
& \simeq \frac{\iu}{8} \,
\Bigl\langle \left(\Xil\Psi\right)^2, \left(\Xl\Psi\right)\Psi \Bigr\rangle
-\frac{\iu}{8} \,
\Bigl\langle \left(\Xil\Psi\right)^2, \Psi\left(\Xl\Psi\right) \Bigr\rangle \,.
\end{split}
\label{quartic-small-lambda}
\end{equation}
Using CFT correlation functions on the unit disk $D_2$,
we can express these terms as
\begin{equation}
\SB_4 \simeq -\frac{\iu}{8} \,
\Big\langle \, \xi (\dc_1) \, \xi (\dc_2) \, X (\dc_3) 
\prod_{k=1}^4 g_k\circ\Psi(0)\Big\rangle_{D_2}
+\frac{\iu}{8} \,
\Big\langle \, \xi (\dc_1) \, \xi (\dc_2) \, X (\dc_4) 
\prod_{k=1}^4 g_k\circ\Psi(0)\Big\rangle_{D_2} \,,
\label{quartic-lambda-to-0}
\end{equation}
where we further dropped terms of $O(\lambda)$
in~(\ref{Xil-Psi-approximation}) and~(\ref{Xl-Psi-approximation}).
When $n=4$, $\dc_k$ in~\eqref{dc_k} for small $\lambda$
can be evaluated as
\begin{equation}
\dc_k=\e^{\frac{\iu\pi}{2}k}\left(\frac{\lambda}{2}\right)^\frac{1}{2}+O(\lambda^\frac{5}{2})\,.
\end{equation}
Using the OPE of $X$ and $\xi$ in~\eqref{X-xi-OPE},
the leading behavior of~$\SB_4$ in the limit $\lambda \to 0$ is given by
\begin{align}
& -\frac{\iu}{8} \,
\Big\langle \, \xi (\dc_1) \, \xi (\dc_2) \, X (\dc_3) 
\prod_{k=1}^4 g_k\circ\Psi(0)\Big\rangle_{D_2}
+\frac{\iu}{8} \,
\Big\langle \, \xi (\dc_1) \, \xi (\dc_2) \, X (\dc_4) 
\prod_{k=1}^4 g_k\circ\Psi(0)\Big\rangle_{D_2} \nonumber \\
& = \frac{\iu}{8} \, \biggl[ \,
-\frac{2}{(\dc_1-\dc_3)^2}+\frac{2}{(\dc_2-\dc_3)^2}
+\frac{2}{(\dc_1-\dc_4)^2}-\frac{2}{(\dc_2-\dc_4)^2} \, \biggr] \,
\Big\langle \xi b\e^{2\phi}(0)
\prod_{k=1}^4g_k\circ\Psi(0) \, \Big\rangle_{D_2} 
+O\bigl(\lambda^{-\frac{1}{2}}\bigr) \nonumber \\
& = -\frac{1}{2\lambda}\,\Big\langle \, \xi b\e^{2\phi}(0)\prod_{k=1}^4g_k\circ\Psi(0) \, \Big\rangle_{D_2}
+O\bigl(\lambda^{-\frac{1}{2}}\bigr)\,.
\label{quartic-leading}
\end{align}
We thus conclude that the quartic interaction~$\SB_4$ in~(\ref{S2})
diverges in the limit $\lambda \to 0$.
We will see in section~\ref{section:regularization} that
divergences of four-point amplitudes in the Witten formulation
are canceled by incorporating this quartic interaction.
We will also see in section~\ref{sec:relation} that
divergences in the gauge variation of the action
for the Witten formulation  at $O(g^2)$
are canceled by incorporating this quartic interaction.
However, we should emphasize here that
the cancellation of these divergences is necessary but is not sufficient.
For the four-point amplitudes, the results in the world-sheet theory
must be precisely reproduced without disagreement in finite terms.
The gauge variation of the action must vanish
for each order in the coupling constant
and no finite terms are allowed to remain for the gauge invariance.
The important point of our approach
is that it enables us to discuss these aspects in a well-defined setting.
In particular, in view of our motivation
to find the relation to the covering of the supermoduli space
of super-Riemann surfaces, it would be important
to understand how the singularity in the Witten formulation is resolved.
As we will see, the detailed structure
of the regularized interactions~\eqref{cubic-regularized}
and~\eqref{quartic-regularized}  turns out to be important,
and it is lost in the evaluation using the OPE in~\eqref{quartic-leading}.

Finally, let us estimate the behavior of higher-point interactions
in the limit $\lambda \to 0$.
The operator $\mathcal{O}$ with the lowest weight which could appear
in the OPE of the operators near the origin in~\eqref{n-point_first_correlation} is
\begin{equation}
\mathcal{O} = b\e^{2\phi}\xi\partial\xi\partial^2\xi\cdots\partial^{n-4}\xi \,,
\end{equation}
and its conformal weight $h_{\mathcal{O}}$ is
\begin{equation}
h_{\mathcal{O}} = -2+\frac{1}{2}\, (n-3)(n-4) \,.
\end{equation}
Since all of the operators near the origin in~\eqref{n-point_first_correlation}
are primary fields of weight 0,
the coefficient in front of $\mathcal{O}$ in the OPE
should compensate the conformal weight of $\mathcal{O}$.
The coefficient is made of $\dc_k$ in~\eqref{dc_k},
and $\dc_k$  for $n$-point interactions
is of $O(\lambda^{2/n})$ for small $\lambda$.
Therefore, we obtain a lower bound $\Delta(n)$ for the power of $\lambda$
in the limit $\lambda \to 0$ for $n$-point interactions given by
\begin{equation}
\Delta(n)=-\frac{4}{n}+\frac{(n-4)(n-3)}{n}\,.
\end{equation}
Since $\Delta(n) > 0$ for $n\geq 6$,
we conclude that $n$-point interactions with $n\geq 6$ vanish in the limit $\lambda \to 0$.
For $n=5$, we find that $\Delta(5) = -2/5$, but it turns out
that singular terms of $O(\lambda^{-2/5})$ vanish
because of a discrete rotational symmetry.
Detailed calculations show that the limit of the five-point interactions is finite.
See~\cite{GIN} for details.

As we discussed before, it is not crucial for our purposes
whether the $n$-point interactions are
divergent, finite, or vanishing in the limit $\lambda \to 0$.
In particular, even though the $n$-point interactions with $n \ge 6$ vanish
in the limit, they may give nonvanishing contributions when they are used
in Feynman diagrams.
If it happens to be the case that higher-point interactions
do not contribute in Feynman diagrams when we take the limit $\lambda \to 0$,
it would be practically useful
and it would be important to understand why it is the case
in the context of the covering of the supermoduli space
of super-Riemann surfaces.

\section{On-shell four-point amplitudes}
\label{section:regularization}
\setcounter{equation}{0} 
In the Witten formulation of open superstring field theory,
on-shell four-point amplitudes at the tree level
suffer from divergences originated from the collision of picture-changing operators.
In this section we investigate how the divergence in the Witten formulation
is resolved in our approach using the Berkovits formulation as a regularization.
In particular, we elucidate the role of the quartic interaction in the Berkovits formulation.
This is one of the main results in this paper.

In subsections~\ref{subsec:covering} and~\ref{sec:bosonic-string-field-theory}
we begin by reviewing how on-shell four-point amplitudes
in the world-sheet theory of the bosonic string are reproduced
in open bosonic string field theory.
Two important points are the covering of the moduli space
of disks with four punctures on the boundary
and the decoupling of BRST-exact states.
In subsection~\ref{subsec:super_decomposition}
we move on to on-shell four-point amplitudes
in the world-sheet theory of the superstring in the RNS formalism.
A new ingredient is the assignment of picture numbers to external states.
In subsection~\ref{subsec:on-shell_berkovits}
we show that the on-shell four-point amplitudes
in the world-sheet theory are correctly reproduced
in open superstring field theory in the Berkovits formulation.
In subsection~\ref{Relation_to_Witten} we use the Berkovits formulation
as a regularization of the Witten formulation
and see how the divergence in the Witten formulation is resolved.

It was shown in~\cite{Berkovits:1999bs} that
open superstring field theory in the Berkovits formulation reproduces
on-shell four-point amplitudes in the world-sheet theory correctly.
We generalize the calculation
of~\cite{Berkovits:1999bs} in such a way that
the relation to the formulation in the small Hilbert space
is seen more clearly.

\subsection{The world-sheet theory in the bosonic string}
\label{subsec:covering}
In the world-sheet theory,
we can calculate on-shell four-point amplitudes at the tree-level
using three unintegrated vertex operators and one integrated vertex operator.
The unintegrated vertex operator
$\Psi(t)$ in the bosonic string is given by
\begin{align}
\label{unintegrated_bosonic}
\Psi(t)=cV_1(t)\,,
\end{align}
where $V_1$ is a conformal
primary field in the matter sector of weight $1$. 
The vertex operator $\Psi(t)$ is BRST closed:
\begin{align}
Q\cdot\Psi(t)=Q\cdot cV_1(t)=0\,.
\end{align}
The integrated vertex operator takes the form
\begin{align}
\int 
\diff t 
\, V_1(t) \,. 
\end{align}
It is BRST invariant up to possible contributions from surface terms
because $Q\cdot V_1(t)$ is a total derivative in $t$:
\begin{align}
Q\cdot V_1(t)
=\partial_t \, [ \, cV_1 (t) \, ]
=\partial_t\Psi(t)\,.
\end{align}
The four-point amplitude $\mathcal{A}_{\rm ws}$
of external states labeled by $A$, $B$, $C$, and $D$
is given by correlation functions of these vertex operators
on the upper half-plane (UHP) as\footnote{
Rigorously speaking, the point $t = \infty$
is outside 
the upper half-plane
and we need another coordinate patch.
However, the operator we insert at $t =\infty$ is always
a primary field of weight $0$
so that we can simply take the limit $t \to \infty$ of $\Psi (t)$.
}
\begin{align}
\label{A_ws_bosonic}
\mathcal{A}_{\rm ws}&=
g^2\int_{-\infty}^\infty 
\diff t \,
\Big\langle \Psi_A(0)
\,V_{1,B}(t)
\,\Psi_C(1)
\,\Psi_D(\infty)
\Big\rangle_{\rm UHP}
+(C\leftrightarrow D)\,. 
\end{align}

Let us decompose
the moduli integral in (\ref{A_ws_bosonic})
with respect to the cyclic ordering
of external states.
For example,
the cyclic ordering
$[ \, A,B,C,D \, ]$
is obtained from
the integral region $0<t<1$
of the first term:
\begin{align}
\label{ws_ABCD_bosonic}
\mathcal{A}^{\rm ws}_{ABCD} 
&=g^2\int_0^1 
\diff t \,
\Big\langle \Psi_A(0)
\,V_{1,B}(t)
\,\Psi_C(1)
\,\Psi_D(\infty)
\Big\rangle_{\rm UHP} \,.
\end{align}
We call amplitudes with a definite cyclic ordering
``color-ordered amplitudes''
by analogy with non-Abelian gauge theories,
and the cyclic ordering of the color-ordered amplitude
is labeled by its subscript as in~\eqref{ws_ABCD_bosonic}.
The amplitude (\ref{A_ws_bosonic}) is then decomposed as
\begin{align}
\mathcal{A}_{\rm ws}
&=
\mathcal{A}^{\rm ws}_{ABCD}
+\mathcal{A}^{\rm ws}_{ABDC}
+\mathcal{A}^{\rm ws}_{ACBD}
+\mathcal{A}^{\rm ws}_{ACDB}
+\mathcal{A}^{\rm ws}_{ADBC}
+\mathcal{A}^{\rm ws}_{ADCB}\,.
\end{align}

\subsection{String field theory in the bosonic string}
\label{sec:bosonic-string-field-theory} 
In string field theory, scattering amplitudes are calculated
in terms of Feynman diagrams just as in ordinary field theory.
The action of open bosonic string field theory is
\begin{equation}
S = -\frac{1}{2} \, \langle \, \Psi, Q \Psi \, \rangle
-\frac{g}{3} \, \langle \, \Psi, \Psi \ast \Psi \, \rangle \,, 
\end{equation}
where $\Psi$ is the open string field of ghost number $1$
and $g$ is the coupling constant.
We need to choose a gauge for calculations of scattering amplitudes
in perturbation theory.
We impose the Siegel-gauge condition given by
\begin{align}
b_0\Psi=0\,,
\end{align}
where $b_0$ is the zero mode of the $b$ ghost.
The propagator $\mathcal{P}$ 
has to
satisfy
\begin{align}
\mathcal{P}Q\Psi=\Psi\,,
\quad
\Psi^\star Q\mathcal{P}=\Psi^\star
\end{align}
for any $\Psi$ satisfying $b_0\Psi=0$, where $\Psi^\star$ is the BPZ conjugate of $\Psi$.
The explicit form of $\mathcal{P}$ is
\begin{align}
\label{Siegel_gauge_propagator}
\mathcal{P}=\frac{b_0}{L_0} \,, 
\end{align}
where $L_0$ is the zero mode of the energy-momentum tensor,
and $1/L_0$ is defined by
\begin{align}
\frac{1}{L_0}=\int_0^\infty 
\diff s \, \e^{-sL_0}\,.
\end{align}

Feynman diagrams for four-point amplitudes at the tree level
consist of two cubic vertices and one propagator.
Just as we did for amplitudes in the world-sheet theory,
let us decompose the four-point amplitude $\mathcal{A}$ in string field theory
into color-ordered amplitudes as follows:
\begin{align}
\mathcal{A}&=\mathcal{A}_{ABCD}
+\mathcal{A}_{ABDC}
+\mathcal{A}_{ACBD}
+\mathcal{A}_{ACDB}
+\mathcal{A}_{ADBC}
+\mathcal{A}_{ADCB} \,,
\end{align}
where $\mathcal{A}_{ABCD}$, for example,
is the color-ordered amplitude with the cyclic ordering
$[ \, A,B,C,D \, ]$
of external states.
The color-ordered amplitude $\mathcal{A}_{ABCD}$
consists of the following two terms:
\begin{align}
\label{ABCD_bosonicSFT}
\mathcal{A}_{ABCD}&
= g^2 \, \Big\langle \Psi_A\ast \Psi_B\,,\frac{b_0}{L_0} \,
(\Psi_C\ast \Psi_D)\Big\rangle
+g^2 \, \Big\langle \Psi_B\ast \Psi_C\,,\frac{b_0}{L_0} \,
(\Psi_D\ast \Psi_A)\Big\rangle \,. 
\end{align}
Let us treat $\Psi_A$ and $\Psi_B$ as incoming states
and $\Psi_C$ and $\Psi_D$ as outgoing states.
Then the first term corresponds to the $s$-channel diagram
and the second term corresponds to the $t$-channel diagram.

The relation between the four-point amplitude
of open bosonic string field theory
in Siegel gauge
and the amplitude in the world-sheet theory is well understood~\cite{Giddings:1986iy}. 
When we map the state $\Psi_A$
to an unintegrated vertex operator at $t=0$,
the state $\Psi_C$
to an unintegrated vertex operator at $t=1$,
and the state $\Psi_D$
to an unintegrated vertex operator at $t=\infty$,
the state $\Psi_B$ is mapped to an integrated vertex operator.
The $s$-channel contribution in~\eqref{ABCD_bosonicSFT}
then corresponds to the region $0 \le t \le 1/2$
of the moduli space in~\eqref{ws_ABCD_bosonic},
and the $t$-channel contribution in~\eqref{ABCD_bosonicSFT}
corresponds to the region $1/2 \le t \le 1$:
\begin{align}
\label{s_ws_bosonic}
\Big\langle \Psi_A\ast \Psi_B\,,\frac{b_0}{L_0} \,
(\Psi_C\ast \Psi_D)\Big\rangle
&=\int_0^{1/2} 
\diff t \, 
\Big\langle \Psi_A(0)
\,V_{1,B}(t)
\,\Psi_C(1)
\,\Psi_D(\infty)
\Big\rangle_{\rm UHP}\,,\\
\label{t_ws_bosonic}
\Big\langle  \Psi_B\ast \Psi_C\,,\frac{b_0}{L_0} \,
(\Psi_D\ast \Psi_A)\Big\rangle&=\int_{1/2}^1 
\diff t \, 
\Big\langle \Psi_A(0)
\,V_{1,B}(t)
\,\Psi_C(1)
\,\Psi_D(\infty)
\Big\rangle_{\rm UHP}\,.
\end{align}
Since the sum of the $s$-channel contribution and the $t$-channel contribution
precisely covers the moduli space $0 \le t \le 1$ in~\eqref{ws_ABCD_bosonic},
the on-shell four-point amplitude in the world-sheet theory
is correctly reproduced in open bosonic string field theory.
The transition from
the $s$ channel~\eqref{s_ws_bosonic}
to the $t$ channel~\eqref{t_ws_bosonic}
occurs at the point $t=1/2$,
which can be regarded as
a boundary between the moduli space covered by the $s$ channel
and that covered by the $t$ channel,
and this boundary plays an important role later.
While the amplitude~(\ref{A_ws_bosonic}) in the world-sheet theory
is expressed in terms of three unintegrated vertex operators
and one integrated vertex operator,
ingredients of the amplitude in string field theory
can be thought of as four unintegrated vertex operators.
The moduli integral of the location of the integrated vertex operator
in the world-sheet theory is transformed
to the integral over the Schwinger parameter $s$
of the propagator in string field theory,
and the associated $b$ ghost is inserted.
The moduli space of disks
with four punctures on the boundary
is covered by Feynman diagrams with cubic vertices alone,
and four-point vertices are not necessary in open bosonic string field theory.

Let us next consider the decoupling of BRST-exact states.
Since two on-shell states which differ by a BRST-exact state
represent the same physical state,
the amplitude in string field theory must vanish
when one of the external states is BRST exact.
Suppose that $\Psi_A$ is BRST exact
and takes the form $\Psi_A=Q\Lambda$:
\begin{align}
\label{ABCD_decoupling_bosonic}
\mathcal{A}_{ABCD}
& = g^2 \, \Big\langle Q\Lambda\ast \Psi_B\,,\frac{b_0}{L_0} \,
(\Psi_C\ast \Psi_D)\Big\rangle
+g^2 \, \Big\langle  \Psi_B\ast \Psi_C\,,\frac{b_0}{L_0} \,
(\Psi_D\ast Q\Lambda)\Big\rangle\,.
\end{align}
Using the properties of the BRST operator
\begin{equation}
\langle \, A, Q B \, \rangle = -(-1)^A \langle \, QA, B \, \rangle \,, \quad
Q \, ( A \ast B ) = Q A \ast B +(-1)^A A \ast Q B
\end{equation}
for any states $A$ 
and
$B$ and the on-shell conditions
$Q \Psi_B = 0$, 
$Q \Psi_C = 0$, 
and $Q \Psi_D = 0$, 
we can rewrite the $s$-channel contribution as
\begin{equation}
\Big\langle Q\Lambda\ast \Psi_B\,,\frac{b_0}{L_0} \,
(\Psi_C\ast \Psi_D)\Big\rangle
= \Big\langle \Lambda\ast \Psi_B\,,
\left\{Q,\frac{b_0}{L_0}\right\}
(\Psi_C\ast \Psi_D)\Big\rangle \,.
\end{equation}
The anticommutator of the BRST operator and the propagator
can be evaluated as
\begin{equation}
\Big\{Q,\frac{b_0}{L_0}\Big\}
=\int_0^\infty 
\diff s \, 
\e^{-sL_0}
\{Q,b_0\}
=\int_0^\infty 
\diff s \, 
\e^{-sL_0}
L_0
= -\int_0^\infty \diff s \, \partial_s \, [ \, \e^{-sL_0} \, ]
= -\Bigl[ \, \e^{-sL_0} \, \Bigr]^{s=\infty}_{s=0} = 1\,,
\label{Q-and-P} 
\end{equation}
and we find
\begin{align}
\Big\langle Q\Lambda\ast \Psi_B\,,\frac{b_0}{L_0} \,
(\Psi_C\ast \Psi_D)\Big\rangle
=\Big\langle \Lambda\ast \Psi_B\,,
\Psi_C\ast \Psi_D\Big\rangle\,.
\label{BRST_WS_1_bosonic}
\end{align}
As can be seen from~\eqref{Q-and-P},
the nonvanishing contribution comes from the surface term at $s=0$
of the integral over $s$,
and it corresponds to the boundary
between the moduli space covered by the $s$ channel
and that covered by the $t$ channel.
Because of this nonvanishing term from the boundary,
the BRST-exact state does not decouple
in the $s$-channel contribution alone.
Similarly, we can rewrite the $t$-channel contribution as
\begin{align}
\label{BRST_WS_2_bosonic}
\Big\langle  \Psi_B\ast \Psi_C\,,\frac{b_0}{L_0} \,
(\Psi_D\ast Q\Lambda)\Big\rangle
&= {}-\Big\langle  \Psi_B\ast \Psi_C\,,
\Psi_D\ast \Lambda\Big\rangle
= {}-\Big\langle  \Lambda\ast\Psi_B\,,
\Psi_C\ast \Psi_D \Big\rangle\,,
\end{align}
where we used
the cyclicity property
of the BPZ inner product
$\langle \, A\ast B,C\ast D \, \rangle
=\langle \, B\ast C,D\ast A \, \rangle$
when $A$ is Grassmann even.
The $t$-channel contribution is also nonvanishing,
but the sum of the $s$-channel contribution
and the $t$-channel contribution vanishes
and the BRST-exact state is decoupled.
The cancellation of the surface terms from the $s$ channel and the $t$ channel
is crucial for the decoupling of BRST-exact states

\subsection{The world-sheet theory in the superstring}
\label{subsec:super_decomposition}
Let us move on to the world-sheet theory of the superstring in the RNS formalism
and consider on-shell disk amplitudes
of four external open-string states in the NS sector.
As in the case of the bosonic string,
we can use three unintegrated vertex operators and one integrated vertex operator,
but we also need to choose picture numbers of the vertex operators.
For disk amplitudes in the small Hilbert space,
the sum of the picture numbers has to be $-2$.
In the following we use
two unintegrated vertex operators in the $-1$ picture,
one unintegrated vertex operator in the $0$ picture,
and one integrated vertex operator in the $0$ picture.
In the world-sheet theory, it is fine to make a convenient choice
of picture numbers for vertex operators this way,
but the asymmetric treatment of vertex operators
or the explicit use of picture-changing operators
is a source of complication in the context of string field theory.
The Berkovits formulation of open superstring field theory
based on the large Hilbert space, however, avoids this complication
in a clever way at least for disk amplitudes with NS vertex operators.
For more general scattering amplitudes in the RNS formalism,
we need to integrate over the supermoduli space
of super-Riemann surfaces with punctures,
and a simple prescription based on picture-changing operations
does not work in general~\cite{Witten:2012ga,Witten:2012bh,Witten:2013cia,Donagi:2013dua}. 
We would need to contemplate its consequence for string field theory,
and we hope that our work will provide a useful approach to this problem.

The unintegrated vertex operator
$\Psi(t)$ in the $-1$ picture is given by
\begin{align}
\label{vertex_-1picture}
\Psi(t)=-c \e^{-\phi}\widehat{V}_{1/2}(t)\,,
\end{align}
where $\widehat{V}_{1/2}$ is a superconformal
primary field in the matter sector of weight $1/2$.
It is BRST closed: $Q\cdot\Psi(t)=0$.
The unintegrated vertex operator in the $0$ picture is obtained
by colliding the picture-changing operator $X(t)$ with $\Psi(t)$:
\begin{align}
X\Psi(t)
:=\lim_{\epsilon\to0}X(t+\epsilon)\Psi(t)
=cV_1(t)+\eta \e^{\phi}\widehat{V}_{1/2}(t)
\quad
{\rm with}
\quad
V_1(t)=G^\mrm{m}_{-1/2}\cdot\widehat{V}_{1/2}(t)\,, 
\end{align}
where $G_{-1/2}^\mrm{m}$ generates the supersymmetry transformation in the matter sector.
The corresponding integrated vertex operator in the $0$ picture is given by
\begin{align}
\int \diff t \,
V_1(t)\,,
\end{align}
and the BRST transformation of $V_1(t)$
reproduces the derivative
of the unintegrated vertex operator
in the $0$ picture:
\begin{align}
Q\cdot V_1(t)=\partial_t \, [ \, cV_1(t)+\eta \e^{\phi}\widehat{V}_{1/2}(t) \, ]
=\partial_t \, [ \, X\Psi(t) \, ] \,.
\end{align}

With these vertex operators, the on-shell disk amplitude
$\mathcal{A}_{\rm ws}$
of external states labeled by $A$, $B$, $C$, and $D$
is given by correlation functions in the small Hilbert space
on the upper half-plane as
\begin{align}
\label{A_ws}
\mathcal{A}_{\rm ws}&=
g^2\int_{-\infty}^\infty 
\diff t \, 
\Bllangle X\Psi_A(0)
\,V_{1,B}(t)
\,\Psi_C(1)
\,\Psi_D(\infty)
\Brrangle_{\rm UHP}
+(C\leftrightarrow D)\,. 
\end{align}

As we did in the bosonic string,
we introduce the color-ordered amplitude $\mathcal{A}^{\rm ws}_{ABCD}$ 
with the cyclic ordering
$[ \, A,B,C,D \, ]$ 
of external states
as follows:
\begin{align} \label{ws_ABCD_super}
\mathcal{A}^{\rm ws}_{ABCD}&= 
g^2\int_0^1 
\diff t \,
\Bllangle X\Psi_A(0)
\,V_{1,B}(t)
\,\Psi_C(1)
\,\Psi_D(\infty)
\Brrangle_{\rm UHP}\,.
\end{align}
Note that
the vertex operators for the states $A$ and $B$ are in the $0$ picture
and those for $C$ and $D$ are in the $-1$ picture.
As we mentioned in the preceding subsection,
ingredients of four-point amplitudes
in open bosonic string field theory
can be thought of as four unintegrated vertex operators.
The relations~\eqref{s_ws_bosonic} and~\eqref{t_ws_bosonic}
depend only on properties of conformal transformations
of Riemann surfaces with punctures
and the familiar relation
between integrated and unintegrated vertex operators
via the action of the $b$ ghost,
so they can be extended to the amplitudes
in the superstring we are discussing.
We thus find
\begin{align} \label{WS_SFT-like}
\mathcal{A}^{\rm ws}_{ABCD} 
& =g^2 \, \Bllangle X_0\Psi_A\ast X_0\Psi_B\,,\frac{b_0}{L_0} \,
(\Psi_C\ast \Psi_D)\Brrangle
+g^2 \, \Bllangle  X_0\Psi_B\ast \Psi_C\,,\frac{b_0}{L_0} \,
(\Psi_D\ast X_0\Psi_A)\Brrangle\,.
\end{align}
Here $X_0$ is the zero-mode of
the picture-changing operator $X$,  
and the operator $X \Psi (0)$ has been mapped to the state $X_0 \Psi$
by the state-operator correspondence because
\begin{equation}
X \Psi(0)
= \lim_{\epsilon \to 0} X(\epsilon) \, \Psi(0)
=\oint_C \frac{\diff z}{2\pi\iu} \, \frac{X(z)}{z} \,\Psi(0)
= X_0 \cdot \Psi(0) \,,
\end{equation}
where the contour $C$ encircles the origin counterclockwise.
The first term on the right-hand side of~\eqref{WS_SFT-like}
corresponds to the $s$-channel contribution
and the second term corresponds to the $t$-channel contribution.
Note that the picture-changing operators $X_0$ are attached
to the states $A$ and $B$ in both channels.

The decoupling of BRST-exact states works in the amplitude~\eqref{WS_SFT-like}
just as in the bosonic string.
Suppose that $\Psi_A$ is BRST-exact
and takes the form $\Psi_A=Q\Lambda$:
\begin{align}
\mathcal{A}^{\rm ws}_{ABCD} 
& =g^2 \, \Bllangle X_0Q\Lambda\ast X_0\Psi_B\,,\frac{b_0}{L_0} \,
(\Psi_C\ast \Psi_D)\Brrangle
+g^2 \, \Bllangle  X_0\Psi_B\ast \Psi_C\,,\frac{b_0}{L_0} \,
(\Psi_D\ast X_0Q\Lambda)\Brrangle\,.
\end{align}
Each of the $s$-channel contribution and the $t$-channel contribution
is nonvanishing and is given by
\begin{align}
\label{BRST_WS_1}
\Bllangle X_0Q\Lambda\ast X_0\Psi_B\,,\frac{b_0}{L_0} \,
(\Psi_C\ast \Psi_D)\Brrangle
&=\Bllangle X_0\Lambda\ast X_0\Psi_B\,,
\Psi_C\ast \Psi_D\Brrangle\,,\\
\label{BRST_WS_2}
\Bllangle  X_0\Psi_B\ast \Psi_C\,,\frac{b_0}{L_0} \,
(\Psi_D\ast X_0Q\Lambda)\Brrangle
&= {}-\Bllangle  X_0\Lambda\ast X_0\Psi_B\,,
\Psi_C\ast \Psi_D \Brrangle\,,
\end{align}
where we used $\{ Q, X_0 \} = 0$.
However, the sum of the two contributions vanish
and the BRST-exact state is decoupled.
We emphasize that
for the cancellation of~\eqref{BRST_WS_1} and~\eqref{BRST_WS_2},
it is important that the picture-changing operators
are attached to the same external states
in the $s$ channel and in the $t$ channel.

The expression~\eqref{WS_SFT-like} for the color-ordered amplitude
can be generalized in various ways.
First, the locations of the picture-changing operators can be changed.
For example, we have
\begin{equation}
\mathcal{A}^{\rm ws}_{ABCD} 
= g^2 \, \Bllangle X_0\Psi_A\ast \Psi_B\,,\frac{b_0}{L_0} \,
(X_0\Psi_C\ast \Psi_D)\Brrangle
+g^2 \, \Bllangle  \Psi_B\ast X_0\Psi_C\,,\frac{b_0}{L_0} \,
(\Psi_D\ast X_0\Psi_A)\Brrangle\,.
\end{equation}
See appendix~\ref{subsub:more_general} for details.
Note that the picture-changing operators
in the $s$ channel and in the $t$ channel
have to be moved in the same way
so that they are attached to the same external states in both channels.
In fact, this property is necessary
for the amplitude $\mathcal{A}_{\rm ws}$ in~\eqref{A_ws}
to be decomposed into the color-ordered amplitudes as
\begin{equation}
\mathcal{A}_{\rm ws}
= \mathcal{A}^{\rm ws}_{ABCD}
+\mathcal{A}^{\rm ws}_{ABDC}
+\mathcal{A}^{\rm ws}_{ACBD}
+\mathcal{A}^{\rm ws}_{ACDB}
+\mathcal{A}^{\rm ws}_{ADBC}
+\mathcal{A}^{\rm ws}_{ADCB}\,. 
\end{equation}
Second, the operator $X_0$ can be replaced with $\Xl$ defined in~\eqref{calX}.
See again appendix~\ref{subsub:more_general} for details.
The color-ordered amplitude $\mathcal{A}^{\rm ws}_{ABCD}$ can then be written,
for example, as 
\begin{equation}
\label{WS_SFT-like_general2}
\mathcal{A}^{\rm ws}_{ABCD} 
= g^2 \, \Bllangle \Xl\Psi_A\ast \Xl\Psi_B\,,\frac{b_0}{L_0} \,
(\Psi_C\ast\Psi_D)\Brrangle
+g^2 \, \Bllangle  \Xl\Psi_B\ast \Psi_C\,,\frac{b_0}{L_0} \,
(\Psi_D\ast \Xl\Psi_A)\Brrangle
\end{equation}
or as
\begin{equation}
\label{WS_SFT-like_general3}
\mathcal{A}^{\rm ws}_{ABCD}
= g^2 \, \Bllangle \Xl\Psi_A\ast \Psi_B\,,\frac{b_0}{L_0} \,
(\Xl\Psi_C\ast \Psi_D)\Brrangle
+g^2 \, \Bllangle  \Psi_B\ast \Xl\Psi_C\,,\frac{b_0}{L_0} \,
(\Psi_D\ast \Xl\Psi_A)\Brrangle\,.
\end{equation}
The goal of the next subsection is to show that
open superstring field theory in the Berkovits formulation reproduces
this color-ordered amplitude~$\mathcal{A}^{\rm ws}_{ABCD}$ 
in the world-sheet theory.

\subsection{String field theory in the superstring}
\label{subsec:on-shell_berkovits}
Let us begin by reviewing
the general structure of four-point amplitudes at the tree level
in the Berkovits formulation which is independent of a gauge choice.
First, we decompose the four-point amplitude $\mathcal{A}$
with external states
$\Phi_A$, $\Phi_B$, $\Phi_C$, and $\Phi_D$ as
\begin{align}
\mathcal{A}=\mathcal{A}_{ABCD}+\mathcal{A}_{ABDC}+\mathcal{A}_{ACBD}+\mathcal{A}_{ACDB}+\mathcal{A}_{ADBC}+\mathcal{A}_{ADCB}\,,
\end{align}
where $\mathcal{A}_{ABCD}$, for example, is the color-ordered amplitude
with the cyclic ordering
$[ \, A,B,C,D \, ]$ 
of external states.
The action in the Berkovits formulation
contains the cubic interaction~$\SB_3$~\eqref{S1}
and the quartic interaction~$\SB_4$~\eqref{S2}.
As in open bosonic string field theory,
contributions to the color-ordered amplitude $\mathcal{A}_{ABCD}$
from Feynman diagrams with two cubic vertices and one propagator
can be decomposed into $\mathcal{A}_s$ for the $s$ channel
and $\mathcal{A}_t$ for the $t$ channel.
In addition, there are contributions from Feynman diagrams
with the quartic interaction, which we denote by $\mathcal{A}_4$.
The color-ordered amplitude $\mathcal{A}_{ABCD}$ is thus given by
\begin{equation}
\label{A}
\mathcal{A}_{ABCD}=\mathcal{A}_s+\mathcal{A}_t+\mathcal{A}_4 \,.
\end{equation}

For the calculation of $\mathcal{A}_s$ and $\mathcal{A}_t$,
it is convenient to rewrite the cubic interaction $\SB_3$ as
\begin{align}
\SB_3 &=
-\frac{\iu}{6} \,
\Big\langle
\Phi,\{Q\Phi,\eta_0\Phi\}
\Big\rangle
\end{align}
and to use the following cyclicity property:
\begin{equation} \label{cyclicity} 
\Big\langle\Phi_1,(Q\Phi_2)(\eta_0\Phi_3)+(\eta_0\Phi_2)(Q\Phi_3)\Big\rangle
=\Big\langle\Phi_2,(Q\Phi_3)(\eta_0\Phi_1)+(\eta_0\Phi_3)(Q\Phi_1)\Big\rangle\,.
\end{equation} 
We can then rewrite Feynman diagrams for $\mathcal{A}_s$ and $\mathcal{A}_t$
in such a way that
the operators $Q$ and $\eta_0$ act only on the external states
and do not act on the propagator $\mathcal{P}$.
The resulting expressions for $\mathcal{A}_s$ and $\mathcal{A}_t$ are
\begin{align}
\mathcal{A}_s &= -\frac{g^2}{4} \, 
\langle \, Q\Phi_A\ast\eta_0\Phi_B+\eta_0\Phi_A\ast Q\Phi_B\,,\mathcal{P} \,
(Q\Phi_C\ast\eta_0\Phi_D+\eta_0\Phi_C\ast Q\Phi_D) \, \rangle\,,
\label{A_s}\\*
\mathcal{A}_t &= -\frac{g^2}{4} \,
\langle \, Q\Phi_B\ast\eta_0\Phi_C+\eta_0\Phi_B\ast Q\Phi_C\,,\mathcal{P} \,
(Q\Phi_D\ast\eta_0\Phi_A+\eta_0\Phi_D\ast Q\Phi_A) \, \rangle\,.
\label{A_t}
\end{align}

For the calculation of $\mathcal{A}_4$, it is convenient to use the expression
for $\SB_4$ in~\eqref{quartic-useful}:
\begin{equation}
\SB_4 = \frac{\iu}{8} \,
\Bigl\langle \left(Q\Phi\right)\left(\eta_0\Phi\right), \Phi^2 \Bigr\rangle
-\frac{\iu}{8} \,
\Bigl\langle \left(\eta_0\Phi\right)\left(Q\Phi\right), \Phi^2 \Bigr\rangle
+\frac{\iu}{12} \,
\Bigl\langle Q\eta_0\Phi, \Phi^3 \Bigr\rangle \,.
\end{equation}
The last term on the right-hand side does not contribute in $\mathcal{A}_4$
because of the on-shell condition $Q \eta_0\Phi = 0$.
The contribution $\mathcal{A}_4$ is thus given by
\begin{equation}
\mathcal{A}_4=\iu\,\frac{g^2}{8}\sum_{\rm cyclic}
\langle \, Q\Phi_A\ast \eta_0\Phi_B- \eta_0\Phi_A\ast Q\Phi_B\,,
\Phi_C\ast\Phi_D \, \rangle \,, 
\label{A_4_on-shell}
\end{equation}
where the sum is over four cyclic permutations of
$[ \, A,B,C,D \, ] \,$.

Let us now consider a gauge choice.
Since we are interested in the relation to the Witten formulation,
we impose the condition $\Xil \Phi = 0$ we introduced in section~\ref{section:Gauge-fixing}
for the partial gauge fixing.
In the preceding subsection, we wrote the color-ordered amplitude
$\mathcal{A}^{\rm ws}_{ABCD}$ using $b_0 / L_0$.
To reproduce the structure $b_0 / L_0$,
we impose the condition $b_0 \Phi = 0$.
Therefore, our gauge conditions can be stated as\footnote{
It is straightforward to extend the gauge conditions to
\begin{equation}
\mathcal{B}\Phi=0\,, \quad \Xi\Phi=0\,,
\nonumber
\end{equation}
where the first condition is the liner $b$-gauge
condition~\cite{Kiermaier:2007jg}
and $\Xi$ is the operator we considered in subsection~\ref{subsec:idea}.}
\begin{equation}
b_0 \Phi = 0 \,, \qquad \Xil \Phi = 0 \,.
\end{equation}
See~\cite{Kroyter:2012ni,Torii:validity}
for detailed discussions
on gauge fixing in the Berkovits formulation.
Since the kinetic operator
takes the form $\iu Q\eta_0$,
the propagator $\mathcal{P}$
should satisfy
\begin{equation}
\mathcal{P} \iu Q \eta_0\Phi = \Phi,
\quad
\Phi^\star \iu Q \eta_0 \mathcal{P} = \Phi^\star
\end{equation}
for an arbitrary state $\Phi$
satisfying $b_0\Phi=\Xil\Phi=0$,
where $\Phi^\star$ is the BPZ conjugate of $\Phi$.
The explicit form of $\mathcal{P}$ is
\begin{equation}
\label{P_in_b-xi}
\mathcal{P}
=-\iu\,\Xi_\lambda\frac{b_0\eta_0}{L_0}\Xi_\lambda \,,
\end{equation}
which can also be written as
\begin{equation}
\label{P_in_b-xi-2}
\mathcal{P}
=-\iu\,\Xi_\lambda\frac{b_0}{L_0}
+\iu\,\Xi_\lambda\frac{b_0}{L_0}\Xi_\lambda\eta_0 \,.
\end{equation}
When we use this form of $\mathcal{P}$ for $\mathcal{A}_s$ in~\eqref{A_s},
the action of $\eta_0$ in the second term on the right-hand side of~\eqref{P_in_b-xi-2} gives
\begin{equation}
\eta_0 \, (Q\Phi_C\ast\eta_0\Phi_D+\eta_0\Phi_C\ast Q\Phi_D)
= {}-Q \eta_0 \Phi_C\ast\eta_0\Phi_D+\eta_0\Phi_C\ast Q \eta_0 \Phi_D = 0 \,,
\end{equation}
where we used the on-shell conditions $Q \eta_0 \Phi_C = 0$ and $Q \eta_0 \Phi_D = 0$.
Therefore, the second term on the right-hand side of~\eqref{P_in_b-xi-2}
does not contribute, and $\mathcal{A}_s$ in~\eqref{A_s} is given by
\begin{equation}
\mathcal{A}_s
=\iu\,\frac{g^2}{4} \,
\langle \, Q\Phi_A\ast\eta_0\Phi_B+\eta_0\Phi_A\ast Q\Phi_B\,,\Xi_\lambda\frac{b_0}{L_0} \,
(Q\Phi_C\ast\eta_0\Phi_D+\eta_0\Phi_C\ast Q\Phi_D) \, \rangle \,.
\label{A_s_in_B-Xi}
\end{equation}
Similarly, the $t$-channel contribution $\mathcal{A}_t$ can be written as
\begin{align}
\mathcal{A}_t
&=\iu\,\frac{g^2}{4} \,
\langle \, Q\Phi_B\ast\eta_0\Phi_C+\eta_0\Phi_B\ast Q\Phi_C\,,\Xi_\lambda\frac{b_0}{L_0} \,
(Q\Phi_D\ast\eta_0\Phi_A+\eta_0\Phi_D\ast Q\Phi_A) \, \rangle\,.
\label{A_t_in_B-Xi}
\end{align}
As we discussed in section~\ref{section:Gauge-fixing},
the string field $\Phi$ under the partial gauge fixing $\Xil \Phi = 0$
can be written as $\Phi = \Xil \Psi$ with $\Psi$ in the small Hilbert space.
The ghost number of $\Psi$ is $1$, and the picture number of $\Psi$ is $-1$.
The condition $b_0 \Phi = 0$ we further impose on $\Phi$
can be translated into the condition
\begin{equation}
b_0 \Psi = 0
\end{equation}
on $\Psi$. The on-shell condition $Q \eta_0 \Phi = 0$
can also be
translated into the condition
\begin{equation}
Q \Psi = 0
\label{on-shell-Psi}
\end{equation}
on $\Psi$.
These two conditions characterize the state corresponding
to the unintegrated vertex operator~(\ref{vertex_-1picture}) in the $-1$ picture
of the world-sheet theory.
The string fields $\eta_0 \Phi$ and $Q \Phi$ can be written as
\begin{equation}
\eta_0 \Phi = \Psi \,, \qquad Q \Phi = Q \Xil \Psi = \{ Q, \Xil \} \Psi = \Xl \Psi \,,
\end{equation}
where we used the on-shell condition~\eqref{on-shell-Psi}.
Note that both 
$\Psi$ and $\Xl \Psi$ are in the small Hilbert space.
When we write
$\Phi_A = \Xil \Psi_A$, 
$\Phi_B = \Xil \Psi_B$, 
$\Phi_C = \Xil \Psi_C$, 
and $\Phi_D = \Xil \Psi_D$,
the $s$-channel contribution $\mathcal{A}_s$
and the $t$-channel contribution $\mathcal{A}_t$
are therefore given by 
\begin{align}
\nonumber
\mathcal{A}_s&=\iu\,\frac{g^2}{4}\Big\langle \mathcal{X}_\lambda\Psi_A\ast\Psi_B+\Psi_A\ast \mathcal{X}_\lambda\Psi_B\,,\Xi_\lambda
\frac{b_0}{L_0} \,
(\mathcal{X}_\lambda\Psi_C\ast\Psi_D+\Psi_C\ast \mathcal{X}_\lambda\Psi_D)\Big\rangle\\
&=\frac{g^2}{4}\Bllangle \mathcal{X}_\lambda\Psi_A\ast\Psi_B+\Psi_A\ast \mathcal{X}_\lambda\Psi_B\,,
\frac{b_0}{L_0} \,
(\mathcal{X}_\lambda\Psi_C\ast\Psi_D+\Psi_C\ast \mathcal{X}_\lambda\Psi_D)\Brrangle\,,
\label{small_s}\\
\nonumber
\mathcal{A}_t&=\iu\,\frac{g^2}{4}\Big\langle \mathcal{X}_\lambda\Psi_B\ast\Psi_C+\Psi_B\ast \mathcal{X}_\lambda\Psi_C\,,\Xi_\lambda
\frac{b_0}{L_0} \,
(\mathcal{X}_\lambda\Psi_D\ast\Psi_A+\Psi_D\ast \mathcal{X}_\lambda\Psi_A)\Big\rangle\\
&=\frac{g^2}{4}\Bllangle \mathcal{X}_\lambda\Psi_B\ast\Psi_C+\Psi_B\ast \mathcal{X}_\lambda\Psi_C\,,
\frac{b_0}{L_0} \,
(\mathcal{X}_\lambda\Psi_D\ast\Psi_A+\Psi_D\ast \mathcal{X}_\lambda\Psi_A)\Brrangle\,.
\label{small_t} 
\end{align}

In open bosonic string field theory,
we saw in subsection~\ref{sec:bosonic-string-field-theory} that
the sum of $\mathcal{A}_s$ and $\mathcal{A}_t$ precisely covers
the moduli space in the color-ordered amplitude of the world-sheet theory.
Let us look at the sum of $\mathcal{A}_s$ and $\mathcal{A}_t$
in open superstring field theory:
\begin{align}
\nonumber
\mathcal{A}_s+\mathcal{A}_t
&=\frac{g^2}{4}\Bllangle \mathcal{X}_\lambda\Psi_A\ast\Psi_B\,,
\frac{b_0}{L_0} \,
(\mathcal{X}_\lambda\Psi_C\ast\Psi_D)\Brrangle
+\frac{g^2}{4}\Bllangle \Psi_B\ast \mathcal{X}_\lambda\Psi_C\,,
\frac{b_0}{L_0} \,
(\Psi_D\ast \mathcal{X}_\lambda\Psi_A)\Brrangle\\*
\nonumber
&\quad
+\frac{g^2}{4}\Bllangle \Psi_A\ast \mathcal{X}_\lambda\Psi_B\,,
\frac{b_0}{L_0} \,
(\Psi_C\ast \mathcal{X}_\lambda\Psi_D)\Brrangle
+\frac{g^2}{4}\Bllangle \mathcal{X}_\lambda\Psi_B\ast\Psi_C\,,
\frac{b_0}{L_0} \,
(\mathcal{X}_\lambda\Psi_D\ast\Psi_A)\Brrangle\\
\nonumber
&\quad
+\frac{g^2}{4}\Bllangle \mathcal{X}_\lambda\Psi_A\ast\Psi_B\,,
\frac{b_0}{L_0} \,
(\Psi_C\ast\mathcal{X}_\lambda\Psi_D)\Brrangle
+\frac{g^2}{4}\Bllangle \mathcal{X}_\lambda\Psi_B\ast\Psi_C\,,
\frac{b_0}{L_0} \,
(\Psi_D\ast\mathcal{X}_\lambda\Psi_A)\Brrangle\\*
&\quad
+\frac{g^2}{4}\Bllangle \Psi_A\ast \mathcal{X}_\lambda\Psi_B\,,
\frac{b_0}{L_0} \,
(\mathcal{X}_\lambda\Psi_C\ast \Psi_D)\Brrangle
+\frac{g^2}{4}\Bllangle \Psi_B\ast \mathcal{X}_\lambda\Psi_C\,,
\frac{b_0}{L_0} \,
(\mathcal{X}_\lambda\Psi_D\ast \Psi_A)\Brrangle \,.
\label{A_s+A_t} 
\end{align}
In the first line, the two picture-changing operators are attached
to the states $\Psi_A$ and $\Psi_C$
both in the $s$ channel and in the $t$ channel,
and we find that the sum of the two terms in the first line
gives $1/4$ of $\mathcal{A}^{\rm ws}_{ABCD}$: 
\begin{equation}
\frac{g^2}{4}\Bllangle \mathcal{X}_\lambda\Psi_A\ast\Psi_B\,,
\frac{b_0}{L_0} \,
(\mathcal{X}_\lambda\Psi_C\ast\Psi_D)\Brrangle
+\frac{g^2}{4}\Bllangle \Psi_B\ast \mathcal{X}_\lambda\Psi_C\,,
\frac{b_0}{L_0} \,
(\Psi_D\ast \mathcal{X}_\lambda\Psi_A)\Brrangle
= \frac{1}{4} \,
\mathcal{A}^{\rm ws}_{ABCD} \,. 
\end{equation}
Similarly, the sum of the two terms in the second line
also gives $1/4$ of $\mathcal{A}^{\rm ws}_{ABCD}$. 

In the third line of~\eqref{A_s+A_t},
the picture-changing operators are attached
to $\Psi_A$ and $\Psi_D$ in the $s$ channel.
If they were attached to $\Psi_A$ and $\Psi_D$ in the $t$ channel as well,
the sum of the two terms in the third line would have given
$1/4$ of $\mathcal{A}^{\rm ws}_{ABCD}$. 
However, the picture-changing operators are attached
to $\Psi_A$ and $\Psi_B$ in the $t$ channel,
so the sum of the two terms in the third line
does not give
$1/4$ of $\mathcal{A}^{\rm ws}_{ABCD}$: 
\begin{equation}
\frac{g^2}{4}\Bllangle \mathcal{X}_\lambda\Psi_A\ast\Psi_B\,,
\frac{b_0}{L_0} \,
(\Psi_C\ast\mathcal{X}_\lambda\Psi_D)\Brrangle
+\frac{g^2}{4}\Bllangle \mathcal{X}_\lambda\Psi_B\ast\Psi_C\,,
\frac{b_0}{L_0} \,
(\Psi_D\ast\mathcal{X}_\lambda\Psi_A)\Brrangle
= \frac{1}{4} \, \mathcal{A}^{\rm ws}_{ABCD} 
+ \Delta \mathcal{A} \,,
\end{equation}
where the deviation $\Delta \mathcal{A}$ can be written as
\begin{equation}
\Delta \mathcal{A}
= \frac{g^2}{4}\Bllangle \mathcal{X}_\lambda\Psi_B\ast\Psi_C\,,
\frac{b_0}{L_0} \,
(\Psi_D\ast\mathcal{X}_\lambda\Psi_A)\Brrangle
-\frac{g^2}{4}\Bllangle \Psi_B\ast\Psi_C\,,
\frac{b_0}{L_0} \,
(\mathcal{X}_\lambda\Psi_D\ast\mathcal{X}_\lambda\Psi_A)\Brrangle \,.
\end{equation}
This deviation can be interpreted as an exchange of a picture-changing operator
between $\Psi_B$ and $\Psi_D$.
Since the moduli space of disks with four punctures is
covered by Feynman diagrams with cubic vertices in open bosonic string field theory,
we expect that the source of the deviation is localized at a point in the moduli space.
In fact, we can rewrite the deviation $\Delta \mathcal{A}$ as follows:
\begin{equation}
\begin{split}
\Delta \mathcal{A}
& = -\iu \, \frac{g^2}{4}\Big\langle Q \Xil \Psi_B\ast\Psi_C\,,
\frac{b_0}{L_0} \,
(\Xil \Psi_D\ast\mathcal{X}_\lambda\Psi_A)\Big\rangle
-\iu \, \frac{g^2}{4}\Big\langle \Xil \Psi_B\ast\Psi_C\,,
\frac{b_0}{L_0} \,
( Q \Xil \Psi_D\ast\mathcal{X}_\lambda\Psi_A)\Big\rangle \\
& = -\iu \, \frac{g^2}{4}\Big\langle \Xil \Psi_B\ast\Psi_C\,,
\Big\{ Q,\frac{b_0}{L_0} \Big\} \,
(\Xil \Psi_D\ast\mathcal{X}_\lambda\Psi_A)\Big\rangle \\
& = -\iu \, \frac{g^2}{4}\Big\langle \Xil \Psi_B\ast\Psi_C\,,
\Xil \Psi_D\ast\mathcal{X}_\lambda\Psi_A \Big\rangle \,. 
\end{split}
\label{Delta-A} 
\end{equation}
Note that the operators $\Xil$ act on $\Psi_B$ and $\Psi_D$,
between which we wanted to exchange a picture-changing operator.
Similarly, the sum of the two terms in the fourth line of~\eqref{A_s+A_t}
does not give
$1/4$ of $\mathcal{A}^{\rm ws}_{ABCD}$ 
and the deviation is
\begin{equation}
\iu \, \frac{g^2}{4}\Big\langle \Xil \Psi_B\ast \mathcal{X}_\lambda \Psi_C\,,
\Xil \Psi_D\ast\Psi_A\Big\rangle \,. 
\label{deviation-fourth-line} 
\end{equation}
Again, we wanted to exchange a picture-changing operator
between $\Psi_B$ and $\Psi_D$,
and the operators $\Xil$ act on these states.
In total, the sum of $\mathcal{A}_s$ and $\mathcal{A}_t$
deviates from $\mathcal{A}^{\rm ws}_{ABCD}$ 
as follows:
\begin{equation}
\mathcal{A}_s + \mathcal{A}_t
= \mathcal{A}^{\rm ws}_{ABCD} 
-\iu \, \frac{g^2}{4}\Big\langle \Xil \Psi_B\ast\Psi_C\,,
\Xil \Psi_D\ast\mathcal{X}_\lambda\Psi_A\Big\rangle
+\iu \, \frac{g^2}{4}\Big\langle \Xil \Psi_B\ast \mathcal{X}_\lambda \Psi_C\,,
\Xil \Psi_D\ast\Psi_A\Big\rangle \,.
\label{total-deviation}
\end{equation} 

Actually, this deviation from the color-ordered amplitude in the world-sheet theory
is precisely canceled
by the contribution $\mathcal{A}_4$ from Feynman diagrams with the quartic interaction.
Let us first spell out the summation over cyclic permutations in~\eqref{A_4_on-shell}:
\begin{equation}
\begin{split}
& \sum_{\rm cyclic}
\langle \, Q\Phi_A\ast \eta_0\Phi_B- \eta_0\Phi_A\ast Q\Phi_B\,,
\Phi_C\ast\Phi_D \, \rangle \\
& = \langle \, Q\Phi_A\ast \eta_0\Phi_B\,, \Phi_C\ast\Phi_D \, \rangle
-\langle \, \eta_0\Phi_A\ast Q\Phi_B\,, \Phi_C\ast\Phi_D \, \rangle \\
& \quad~ +\langle \, Q\Phi_B\ast \eta_0\Phi_C\,, \Phi_D\ast\Phi_A \, \rangle
-\langle \, \eta_0\Phi_B\ast Q\Phi_C\,, \Phi_D\ast\Phi_A \, \rangle \\
& \quad~ +\langle \, Q\Phi_C\ast \eta_0\Phi_D\,, \Phi_A\ast\Phi_B \, \rangle
-\langle \, \eta_0\Phi_C\ast Q\Phi_D\,, \Phi_A\ast\Phi_B \, \rangle \\
& \quad~ +\langle \, Q\Phi_D\ast \eta_0\Phi_A\,, \Phi_B\ast\Phi_C \, \rangle
-\langle \, \eta_0\Phi_D\ast Q\Phi_A\,, \Phi_B\ast\Phi_C \, \rangle \,.
\label{cyclic-summation-explicitly}
\end{split}
\end{equation}
After the partial gauge fixing, the string fields
$\Phi$, $\eta_0 \Phi$, and $Q \Phi$
are written as
$\Xil \Psi$, $\Psi$, and $\Xl \Psi$, respectively.
We want the operators $\Xil$
to act on $\Psi_B$ and $\Psi_D$.
In fact, we can make the operators $Q$ and $\eta_0$ act only on $\Phi_A$ and $\Phi_C$
by arranging the terms in~\eqref{cyclic-summation-explicitly}: 
\begin{equation}
\begin{split}
& \langle \, Q\Phi_A \,,  \eta_0 \Phi_B \ast \Phi_C \ast \Phi_D
+\Phi_B \ast \Phi_C \ast \eta_0 \Phi_D \, \rangle
-\langle \, \eta_0 \Phi_A \,,  Q \Phi_B \ast \Phi_C \ast \Phi_D
+\Phi_B \ast \Phi_C \ast Q \Phi_D \, \rangle \\
& +\langle \, Q\Phi_C \,,  \eta_0 \Phi_D \ast \Phi_A \ast \Phi_B
+\Phi_D \ast \Phi_A \ast \eta_0 \Phi_B \, \rangle
-\langle \, \eta_0 \Phi_C \,,  Q \Phi_D \ast \Phi_A \ast \Phi_B
+\Phi_D \ast \Phi_A \ast Q \Phi_B \, \rangle \\
& = {}-\langle \, Q\Phi_A \,, \Phi_B \ast \eta_0 \Phi_C \ast \Phi_D \, \rangle
+\langle \, \eta_0 \Phi_A \,, \Phi_B \ast Q \Phi_C \ast \Phi_D \, \rangle \\
& \quad~ -\langle \, Q\Phi_C \,, \Phi_D \ast \eta_0 \Phi_A \ast \Phi_B \, \rangle
+\langle \, \eta_0 \Phi_C \,, \Phi_D \ast Q \Phi_A \ast \Phi_B \, \rangle \\
& = -2 \, \langle \, Q\Phi_A \,, \Phi_B \ast \eta_0 \Phi_C \ast \Phi_D \, \rangle
+2 \, \langle \, \eta_0 \Phi_A \,, \Phi_B \ast Q \Phi_C \ast \Phi_D \, \rangle \,,
\end{split}
\end{equation} 
where we used the on-shell conditions $Q \eta_0 \Psi_A = 0$ and $Q \eta_0 \Psi_C = 0$.
The contribution $\mathcal{A}_4$ now simplifies to
\begin{equation}
\mathcal{A}_4 =
-\iu\,\frac{g^2}{4}
\Bigl\langle Q\Phi_A \,, \Phi_B \ast \eta_0 \Phi_C \ast \Phi_D \Bigr\rangle
+\iu\,\frac{g^2}{4}
\Bigl\langle \eta_0 \Phi_A \,, \Phi_B \ast Q \Phi_C \ast \Phi_D \Bigr\rangle \,.
\end{equation}
In terms of $\Psi_A$, $\Psi_B$, $\Psi_C$, and $\Psi_D$ in our gauge,
this can be written as
\begin{equation}
\mathcal{A}_4= -\iu\,\frac{g^2}{4}
\Bigl\langle \Xl \Psi_A \,, \Xil \Psi_B \ast \Psi_C \ast \Xil \Psi_D \Bigr\rangle
+\iu\,\frac{g^2}{4}
\Bigl\langle \Psi_A \,, \Xil \Psi_B \ast \Xl \Psi_C \ast \Xil \Psi_D \Bigr\rangle \,.
\label{A_4-simplified} 
\end{equation}
This precisely cancels the deviation in~\eqref{total-deviation},
and the Berkovits formulation reproduces the amplitude in the world-sheet theory correctly:
\begin{align}
\mathcal{A}_{ABCD}
&=\mathcal{A}_s+\mathcal{A}_t+\mathcal{A}_4
=\mathcal{A}^{\rm ws}_{ABCD}\,. 
\end{align}
The quartic interaction in the Berkovits formulation
is necessary 
for reproducing the correct on-shell amplitude,
and we believe that we have elucidated its role:
the quartic interaction adjusts the difference
in the assignment of picture-changing operators
in the $s$ channel and in the $t$ channel.

\subsection{Relation to the Witten formulation}
\label{Relation_to_Witten}
Let us now investigate the limit $\lambda \to 0$
of the on-shell four-point amplitudes in the superstring.
First, consider the color-ordered amplitude $\mathcal{A}^{\rm ws}_{ABCD}$
in the world-sheet theory
of the form~\eqref{WS_SFT-like_general3}:
\begin{equation}
\mathcal{A}^{\rm ws}_{ABCD}
=g^2 \, \Bllangle \Xl\Psi_A\ast \Psi_B\,,
\frac{b_0}{L_0} \,
(\Xl\Psi_C\ast \Psi_D)\Brrangle
+g^2 \, \Bllangle  \Psi_B\ast \Xl\Psi_C\,,
\frac{b_0}{L_0} \,
(\Psi_D\ast \Xl\Psi_A)\Brrangle\,.
\label{WS_SFT-like_general3-again}
\end{equation}
As we explained in subsection~\ref{subsec:gfc},
the operator $\Xl$ for small $\lambda$
can be approximated by a local insertion of the picture-changing operator
near the midpoint.
We expect each of the two terms
in~\eqref{WS_SFT-like_general3-again}
to diverge in the limit $\lambda \to 0$
because two picture-changing operators collide
when the Schwinger parameter $s$ of the propagator is small.
By a calculation similar to 
\eqref{general-lambda-s} or \eqref{general-lambda-t} in appendix~\ref{subsub:more_general},
we can extract such divergences.
For the $s$-channel contribution,
we have
\begin{equation}
\label{WS-s-channel-divergence}
\begin{split}
& \Bllangle \Xl \Psi_A \ast \Psi_B, \frac{b_0}{L_0} \,
( \Xl \Psi_C \ast \Psi_D) \Brrangle \\
& = \Bllangle \Xl \Psi_A \ast \Psi_B, \frac{b_0}{L_0} \,
( X_0 \Psi_C \ast \Psi_D) \Brrangle
+ \Bllangle \Xl \Psi_A \ast \Psi_B, \frac{b_0}{L_0} \,
\big[ \, ( \Xl -X_0 ) \, \Psi_C \ast \Psi_D \, \big] \, \Brrangle \\
& = \Bllangle \Xl \Psi_A \ast \Psi_B, \frac{b_0}{L_0} \,
( X_0 \Psi_C \ast \Psi_D) \Brrangle
+ \Bllangle \Xl \Psi_A \ast \Psi_B,\,
( \Xil -\xi_0 ) \, \Psi_C \ast \Psi_D \, \Brrangle \\
& = \Bllangle \Xl \Psi_A \ast \Psi_B, \frac{b_0}{L_0} \,
( X_0 \Psi_C \ast \Psi_D) \Brrangle
+\iu \, \Big\langle \Xl \Psi_A \ast \Psi_B 
\ast ( \Xil -\xi_0 ) \, \Psi_C,\, \xi_0 \Psi_D  \Big\rangle \\
& = \Bllangle \Xl \Psi_A \ast \Psi_B, \frac{b_0}{L_0} \,
( X_0 \Psi_C \ast \Psi_D) \Brrangle
-\iu \, \Big\langle \Xl \Psi_A \ast \Psi_B 
\ast \xi_0 \Psi_C,\, \xi_0 \Psi_D  \Big\rangle \\
& \quad~ +\iu \, \Big\langle \Xl \Psi_A \ast \Psi_B 
\ast \Xil \Psi_C,\, \xi_0 \Psi_D  \Big\rangle \,,
\end{split}
\end{equation}
where the first two terms in the final expression
are finite and the last term is divergent in the limit $\lambda \to 0$
because of the collision of $X$ and $\xi$ near the midpoint.
We know, however, the color-ordered amplitude $\mathcal{A}^{\rm ws}_{ABCD}$
is independent of $\lambda$,
and the divergence in the $s$ channel is canceled by
the same divergence in the $t$ channel, which can be seen as follows:
\begin{equation}
\begin{split}
& \Bllangle \Psi_B \ast \Xl \Psi_C, \frac{b_0}{L_0} \,
( \Psi_D \ast \Xl \Psi_A ) \Brrangle \\
& = \Bllangle \Psi_B \ast X_0 \Psi_C, \frac{b_0}{L_0} \,
( \Psi_D \ast \Xl \Psi_A ) \Brrangle
-\Bllangle \Psi_B \ast ( \Xil -\xi_0 ) \, \Psi_C,\,
\Psi_D \ast \Xl \Psi_A \, \Brrangle \\
& = \Bllangle \Psi_B \ast X_0 \Psi_C, \frac{b_0}{L_0} \,
( \Psi_D \ast \Xl \Psi_A ) \Brrangle
-\Bllangle \Xl \Psi_A \ast \Psi_B,\,
( \Xil -\xi_0 ) \, \Psi_C \ast \Psi_D \, \Brrangle \\
& = \Bllangle \Psi_B \ast X_0 \Psi_C, \frac{b_0}{L_0} \,
( \Psi_D \ast \Xl \Psi_A ) \Brrangle
+\iu \, \Big\langle \Xl \Psi_A \ast \Psi_B 
\ast \xi_0 \Psi_C,\, \xi_0 \Psi_D  \Big\rangle \\
& \quad~ -\iu \, \Big\langle \Xl \Psi_A \ast \Psi_B 
\ast \Xil \Psi_C,\, \xi_0 \Psi_D  \Big\rangle \,.
\end{split}
\end{equation}
It is also interesting to consider the limit $\lambda \to 0$
of the amplitude $\mathcal{A}^{\rm ws}_{ABCD}$
written in the form~\eqref{WS_SFT-like_general2}:
\begin{equation}
\mathcal{A}^{\rm ws}_{ABCD}
= g^2 \, \Bllangle \Xl\Psi_A\ast \Xl\Psi_B\,,\frac{b_0}{L_0} \,
(\Psi_C\ast\Psi_D)\Brrangle
+g^2 \, \Bllangle  \Xl\Psi_B\ast \Psi_C\,,\frac{b_0}{L_0} \,
(\Psi_D\ast \Xl\Psi_A)\Brrangle \,.
\end{equation}
In the $s$ channel, two picture-changing operators collide
in the whole region of the Schwinger parameter.
However, via a manipulation similar to~\eqref{WS-s-channel-divergence}
in the previous case,
the divergence can be localized at $s=0$
and is canceled by the divergence in the $t$ channel.

Let us move on to the color-ordered amplitude $\mathcal{A}_{ABCD}$
of open superstring field theory.
The sum of the $s$-channel contribution $\mathcal{A}_s$
and the $t$-channel contribution $\mathcal{A}_t$ is given in~\eqref{A_s+A_t}.
As is clear from the discussion so far,
each of the eight terms in~\eqref{A_s+A_t} diverges
in the limit $\lambda \to 0$.
In the first line and in the second line of~\eqref{A_s+A_t},
the divergence from the $s$ channel and the divergence from the $t$ channel
cancel.
In the third line, however, the divergence from the $s$ channel
is not canceled by the divergence from the $t$ channel,
and the total divergence in the third line is contained
within $\Delta \mathcal{A}$ given in~\eqref{Delta-A}:
\begin{equation}
\Delta \mathcal{A}
= -\iu \, \frac{g^2}{4}\Big\langle \Xil \Psi_B\ast\Psi_C\,,
\Xil \Psi_D\ast\mathcal{X}_\lambda\Psi_A \Big\rangle \,.
\end{equation}
In the limit $\lambda \to 0$, the deviation $\Delta \mathcal{A}$ diverges
because two insertions of $\xi$ and one insertion of $X$ collide.
We can evaluate the divergence explicitly
by a calculation similar to the one in~\eqref{quartic-leading}.
Similarly, the divergence from the $s$ channel and the divergence from the $t$ channel
do not cancel in the fourth line of~\eqref{A_s+A_t},
and the total divergence is contained in the deviation given in~\eqref{deviation-fourth-line}.

It is often said that the on-shell four-point amplitude in the Witten formulation
is divergent because two picture-changing operators collide.
As we have seen in this subsection, however,
it is possible that the divergence from the $s$ channel
and the divergence from the $t$ channel cancel.
In our regularization scheme based on the partial gauge fixing
of the Berkovits formulation,
the essence of the divergence in the limit $\lambda \to 0$
is the difference between the divergences in the $s$ channel
and in the $t$ channel.
The divergence can be written in a form
localized at the boundary
between the moduli spaces in the $s$ channel and in the $t$ channel,
and it is canceled by the contribution $\mathcal{A}_4$ from Feynman diagrams
with the quartic interaction.

We should emphasize that the role of the quartic interaction
is not only to cancel the divergence from Feynman diagrams
with two cubic vertices.
As we have seen in subsection~\ref{subsec:on-shell_berkovits},
the quartic interaction adjusts the difference
in the assignment of picture-changing operators
in the $s$ channel and in the $t$ channel.
When $\lambda$ is small,
picture-changing operators approximately localize near the midpoint,
and the different assignment of picture-changing operators
metamorphoses into the subtle difference
in how they approach the midpoint.
The quartic interaction precisely compensates this subtle difference,
and the detailed configuration of the local operators
in the quartic interaction~\eqref{quartic-lambda-to-0} is crucial for this compensation.

The divergence of the on-shell four-point amplitude
in the Witten formulation
can also be
regularized
by introducing a cutoff in the integral region of the Schwinger parameter
for the propagator.
We can then add a quartic interaction as a counterterm
to cancel the divergence.
This is the regularization scheme by Wendt~\cite{Wendt},
and this can be interpreted as regularizing
the integral for the bosonic moduli space
of disks with four punctures on the boundary.
In view of our motivation
to obtain insight into the covering of the supermoduli space
of super-Riemann surfaces,
it would be better if we can keep the bosonic direction intact
and focus on the fermionic direction of the supermoduli space.
In our approach, we emphasize that we do not need to regularize
the integral of the Schwinger parameter.
We regard this as an advantage of our approach.
In fact, it would be difficult to see the role
of the quartic interaction discussed in this subsection
if we had regularized the integral of the Schwinger parameter.

\section{Gauge invariance}
\label{sec:relation}
\setcounter{equation}{0}
As we mentioned in subsection~\ref{Witten formulation},
the variation of the action~\eqref{Witten:NS-action}
in the Witten formulation of open superstring field theory
under the gauge transformation~\eqref{singular gauge transf}
is not well defined at $O(g^2)$
because of the collision of picture-changing operators.
In this section, we use our regularization of the Witten formulation 
in terms of the Berkovits formulation under the partial gauge fixing
developed in section~\ref{section:Gauge-fixing}
and show that the gauge invariance at $O(g^2)$ is recovered
by incorporating the quartic interaction which diverges
in the singular limit $\lambda \to 0$ of the regularization parameter.
While it turns out that this requires fairly complicated calculations,
we demonstrate that the idea of using the residual gauge symmetry
after the partial gauge fixing of the Berkovits formulation
leads to a systematic approach to this problem.

\subsection{Regularization}
\label{gauge-transformation-regularized}
As we explained in subsection~\ref{subsec:idea},
the kinetic term $\SB_2$ in the Berkovits formulation
reduces to the kinetic 
term 
in the Witten formulation
under the partial gauge fixing:
\begin{equation} \label{del1:Psi}
\SB_2 = -\frac{1}{2} \, \bllangle\Psi, Q\Psi\brrangle \,.
\end{equation}
Under the gauge transformation
\begin{equation}
\delta^{(1)}\Psi = Q\Lambda\,,
\end{equation}
where the gauge parameter $\Lambda$ is in the small Hilbert space,
the kinetic term $\SB_2$ is invariant:
\begin{equation}
\delta^{(1)} \SB_2 = 0 \,.
\end{equation}
The cubic interaction in the Witten formulation
is regularized as~\eqref{cubic-regularized} in our approach:
\begin{equation}
\SB_3 = \frac{\iu}{6} \, \Big\langle\Psi, \big[ \Xil\Psi, Q\Xil\Psi\big]\Big\rangle \,.
\end{equation}
While this term is written in the language of the large Hilbert space,
we regard it as a term in an action for $\Psi$ in the small Hilbert space.
The combined action $\SB_2 +g \, \SB_3$ is gauge invariant at $O(g)$
if we can construct a nonlinear correction $\delta^{(2)} \Psi$
to the gauge transformation $\delta^{(1)} \Psi$
such that 
\begin{equation}
\delta^{(2)} \SB_2 + \delta^{(1)}\SB_3 =0
\label{gauge-invariance-g^1}
\end{equation}
is satisfied.

Let us construct the correction $\delta^{(2)} \Psi$ to the gauge transformation
satisfying~\eqref{gauge-invariance-g^1}.
It is convenient to transform $\SB_3$ as follows:
\begin{equation}
\SB_3 = \frac{\iu}{6} \, \Big\langle\Psi, \big[ \Xil\Psi, Q\Xil\Psi\big]\Big\rangle
= -\frac{\iu}{6} \, \Bigl\langle\bigl\{\Psi, Q\Xil\Psi\bigr\},\hs \Xil\hs \Psi\Bigr\rangle \,.
\end{equation}
We can then use the cyclicity property~\eqref{cyclicity},
and $\delta^{(1)}\SB_3$ is given by
\begin{equation}
\delta^{(1)}\SB_3 = -\frac{\iu}{2} \,
\Bigl\langle\bigl\{\Psi, Q\Xil\Psi\bigr\},\hs \Xil\hs\delta^{(1)}\Psi\Bigr\rangle
= -\frac{\iu}{2} \,
\Bigl\langle\bigl\{\Psi, Q\Xil\Psi\bigr\},\hs \Xil Q\Lambda\Bigr\rangle\,.
\label{del1:SB3}
\end{equation}
Since $\delta^{(2)} \SB_2$ takes the form
\begin{equation}
\delta^{(2)}\SB_2 = -\bllangle Q\Psi, \delta^{(2)}\Psi \brrangle\,,
\label{del2:SB2}
\end{equation}
our goal is to rewrite $\delta^{(1)}\SB_3$
in a form of a BPZ inner product of $Q \Psi$ and a state in the small Hilbert space.
Let us transform $\delta^{(1)}\SB_3$ in the following way:
\begin{align}
\delta^{(1)}\SB_3 &=
-\frac{\iu}{2}\Bigl\langle\Psi, \bigl[Q\Xil\Psi, \Xil Q\Lambda\bigr]\Bigr\rangle
= -\frac{\iu}{2}\Bigl\langle\ez\Xil\Psi, \bigl[Q\Xil\Psi, \Xil Q\Lambda\bigr]\Bigr\rangle
\nonumber \\[1ex]
&= \frac{\iu}{2}\Bigl\langle\Xil\Psi, \ez\bigl[Q\Xil\Psi, \Xil Q\Lambda\bigr]\Bigr\rangle
= -\frac{\iu}{2}\Bigl\langle\Xil\Psi, \bigl[Q\Psi, \Xil Q\Lambda\bigr]\Bigr\rangle
-\frac{\iu}{2}\Bigl\langle\Xil\Psi, \bigl\{ Q\Xil\Psi, Q\Lambda\bigr\}\Bigr\rangle 
\nonumber \\[1ex]
&= \frac{\iu}{2}\Bigl\langle Q\Psi, \bigl[\Xil\Psi, \Xil Q\Lambda\bigr]\Bigr\rangle
+\frac{\iu}{2} \Bigl\langle Q\Lambda, \bigl[\Xil\Psi, Q\Xil\Psi\bigr]\Bigr\rangle
\nonumber \\[2ex]
&= \frac{\iu}{2}\Bigl\langle \ez\Xil Q\Psi, \bigl[\Xil\Psi, \Xil Q\Lambda\bigr]\Bigr\rangle
+\frac{\iu}{2} \Bigl\langle \ez\Xil Q\Lambda, \bigl[\Xil\Psi, Q\Xil\Psi\bigr]\Bigr\rangle
\nonumber \\[2ex]
&= \frac{\iu}{2}\Bigl\langle \Xil Q\Psi, \ez\bigl[\Xil\Psi, \Xil Q\Lambda\bigr]\Bigr\rangle
-\frac{\iu}{2} \Bigl\langle \Xil Q\Lambda, \ez\bigl[\Xil\Psi, Q\Xil\Psi\bigr]\Bigr\rangle
\nonumber \\[1ex]
&=\frac{1}{2}\Bllangle Q\Psi,\, \bigl[\Psi, \Xil Q\Lambda\bigr] + \bigl[\Xil\Psi, Q\Lambda\bigr]\Brrangle
-\frac{\iu}{2} \Bigl\langle \Xil Q\Lambda, \ez\bigl[\Xil\Psi, Q\Xil\Psi\bigr]\Bigr\rangle\,.
\label{deform-1}
\end{align}
The first term in the last line takes a form of a BPZ inner product
of $Q \Psi$ and a state in the small Hilbert space.
The second term can be further transformed as follows:
\begin{align}
&-\frac{\iu}{2} \Bigl\langle \Xil Q\Lambda, \ez\bigl[\Xil\Psi, Q\Xil\Psi\bigr]\Bigr\rangle
= -\frac{\iu}{2} \Bigl\langle \Xil Q\Lambda,\, \bigl\{\Psi, Q\Xil\Psi\bigr\} - \bigl[\Xil\Psi, Q\Psi\bigr]\Bigr\rangle
\nonumber \\*[1ex]
=\ &\frac{\iu}{2}\Bigl\langle \Xil Q\Lambda,\, Q\bigl[\Psi, \Xil\Psi\bigr] + 2\bigl[\Xil\Psi, Q\Psi\bigr]\Bigr\rangle
=\frac{\iu}{2} \Bigl\langle \Xl\Lambda, Q\bigl[\Psi, \Xil\Psi\bigr] \Bigr\rangle
-\iu\hs \Bigl\langle Q\Psi, \bigl[  \Xil\Psi, \Xil Q\Lambda\bigr]\Bigr\rangle 
\nonumber \\[1ex]
=\ &\frac{1}{2}\Bllangle \Xl\Lambda, Q\bigl\{\Psi, \Psi\bigr\}\Brrangle
-\iu\hs \Bigl\langle \ez\Xil Q\Psi, \bigl[  \Xil\Psi, \Xil Q\Lambda\bigr]\Bigr\rangle
\nonumber \\[1ex]
=\ &\Bllangle Q\Psi, \bigl[\Psi, \Xl\Lambda\bigr]\Brrangle
-\iu\hs \Bigl\langle \Xil Q\Psi, \ez\bigl[  \Xil\Psi, \Xil Q\Lambda\bigr]\Bigr\rangle
\nonumber \\[1ex]
=\ &\Bllangle Q\Psi,\, \bigl[\Psi, \Xl\Lambda\bigr] -\bigl[\Psi, \Xil Q\Lambda\bigr]- \bigl[\Xil\Psi, Q\Lambda\bigr]\Brrangle\,.
\label{1st-term}
\end{align} 
Therefore, $\delta^{(1)}\SB_3$ can be rewritten as
\begin{equation} \label{del^1}
\delta^{(1)}\SB_3
= \Bllangle Q\Psi,\, \bigl[\Psi, \Xl\Lambda\bigr] -\frac{1}{2}\bigl[\Psi, \Xil Q\Lambda\bigr]- \frac{1}{2}\bigl[\Xil\Psi, Q\Lambda\bigr]\Brrangle\,,
\end{equation}
and $\delta^{(2)}\Psi$ satisfying~\eqref{gauge-invariance-g^1} is given by
\begin{equation}
\delta^{(2)}\Psi
= \bigl[\Psi, \Xl\Lambda\bigr]
-\frac{1}{2}\bigl[\Psi, \Xil Q\Lambda\bigr]
-\frac{1}{2}\bigl[\Xil\Psi, Q\Lambda\bigr] \,.
\label{delta^(2)-Psi}
\end{equation}
In the limit $\lambda \to 0$, the operators $\Xl$ and $\Xil$
become the midpoint insertions $X_{\rm mid}$ and $\xi_{\rm mid}$, respectively.
The limit of $\delta^{(2)}\Psi$ is
\begin{equation}
\begin{split}
\lim_{\lambda \to 0} \delta^{(2)} \Psi
& = \bigl[\Psi, X_{\rm mid}\Lambda\bigr]
-\frac{1}{2} \, \bigl[\Psi, \xi_{\rm mid} Q\Lambda\bigr]
-\frac{1}{2} \, \bigl[\xi_{\rm mid} \Psi,Q\Lambda\bigr] \\
& = X_{\rm mid} \bigl[\Psi, \Lambda\bigr]
+\frac{1}{2} \, \xi_{\rm mid} \bigl\{ \Psi, Q\Lambda\bigr\}
-\frac{1}{2} \, \xi_{\rm mid} \bigl\{  \Psi,Q\Lambda\bigr\} \\
& = X_{\rm mid} \bigl[\Psi, \Lambda\bigr] \,,
\end{split}
\end{equation}
where we used~\eqref{X_mid-identities} and
\begin{equation}
( \, \xi_{\rm mid} A \, ) \, B = (-1)^A \, A \, ( \, \xi_{\rm mid} B \, )
= \xi_{\rm mid} ( A B \, )
\label{xi_mid-identities}
\end{equation}
for any pair of string fields $A$ and $B$.
As expected, $\delta^{(2)}\Psi$ in the limit $\lambda \to 0$
reproduces the nonlinear term
of the gauge transformation~\eqref{singular gauge transf} in the Witten formulation,
and $\delta^{(2)}\Psi$ with finite $\lambda$ provides its regularization.

Now the gauge variation $\delta^{(2)} \SB_3$ at $O(g^2)$ is well defined
for finite $\lambda$.
If it vanishes in the limit $\lambda \to 0$,
we conclude that the action
defined by the limit $\lambda \to 0$ of $\SB_2 +g \, \SB_3$
is gauge invariant.
In case it is nonvanishing in the limit,
the action $\SB_2 +g \, \SB_3$ can still be gauge invariant at $O(g^2)$
if we can construct a correction $\delta^{(3)} \Psi$ to the gauge transformation
such that $\delta^{(2)} \SB_3 + \delta^{(3)}\SB_2 =0$ is satisfied.
If the gauge variation $\delta^{(2)} \SB_3$ is nonvanishing
and cannot be absorbed by correcting the gauge transformation,
we conclude that the gauge invariance is violated at $O(g^2)$.
However, the gauge invariance at $O(g^2)$ can be recovered
by adding a term $g^2 \, \SB_4$ to the action
if $\delta^{(1)} \SB_4 + \delta^{(2)} \SB_3 + \delta^{(3)} \SB_2 = 0$
is satisfied for an appropriate correction term $\delta^{(3)} \Psi$
to the gauge transformation.
Since we have a well-defined expression of $\delta^{(2)} \SB_3$
when $\lambda$ is finite, we can start investigating the gauge invariance
at $O(g^2)$ this way.
It turns out, however, that the expression of $\delta^{(2)} \SB_3$ is fairly complicated.

Actually, the Berkovits formulation provides an answer to the question we are asking.
We constructed a regularized theory based on the partial gauge fixing
of the Berkovits formulation.
The residual gauge symmetry of the Berkovits formulation 
after the partial gauge fixing
guarantees the gauge invariance of the regularized theory,
and we can systematically construct corrections to the gauge transformation
in the expansion with respect to $g$ from the original gauge transformation
in the Berkovits formulation.
We will develop this method in the next subsection.

\subsection{The residual gauge symmetry}
\label{res:gauge}
Let us begin with the free theory.
The action in the Berkovits formulation is invariant under the gauge transformation
\begin{equation}
\delta \Phi = Q \LQ + \ez \Leta \,,
\label{gauge-transformation-free}
\end{equation}
where $\LQ$ and $\Leta$ are the gauge parameters.
We impose
\begin{equation}
\Xil \Phi = 0
\end{equation}
on $\Phi$ for 
partial gauge fixing, and we write $\Phi$ satisfying this condition as
\begin{equation}
\Phi = \Xil \Psi
\end{equation}
with $\Psi$ in the small Hilbert space.
The variation $\delta \Phi$ keeping this condition has to satisfy
\begin{equation}
\Xil \, \delta \Phi
= \Xil ( \, Q \LQ + \ez \Leta \, ) = 0 \,.
\label{constraint-free}
\end{equation}
The parameters $\LQ$ and $\Leta$ are related by this constraint.
We can solve this constraint by acting with $\ez$ on~\eqref{constraint-free}:
\begin{equation}
\ez \Xil \, \delta \Phi
= \ez \Xil ( \, Q \LQ + \ez \Leta \, )
= \ez \Xil Q \LQ + \ez \Xil \ez \Leta
= \ez \Xil Q \LQ + \ez \Leta= 0 \,.
\end{equation}
We find that $\ez\Leta$ can be expressed in terms of $\LQ$ as follows:
\begin{equation}
\ez \Leta = -\ez \Xil Q \LQ \,.
\end{equation}
The variation satisfying this relation can be written as
\begin{equation}
\delta \Phi = Q \LQ + \ez \Leta
= Q \LQ -\ez \Xil Q \LQ
= ( 1-\ez \Xil ) \, Q \LQ
= \Xil \ez Q \LQ \,.
\end{equation}
This takes the form of $\delta \Phi = \Xil \delta \Psi$ with $\delta \Psi$ given by
\begin{equation}
\delta \Psi = \ez Q \LQ \,. 
\end{equation}
The gauge parameter $\LQ$ is in the large Hilbert space
and can be decomposed as $\LQ = A +\Xil B$,
where $A$ and $B$ are in the small Hilbert space
and given by $A = \ez \Xil \LQ$ and $B = \ez \LQ$.
The term $A$ does not contribute in the variation $\delta \Psi$
because $\ez Q A = -Q \ez A = 0$.
The nonvanishing part of the variation
is then $\ez Q \Xil B = -Q \ez \Xil B = -Q B$,
and we identify it with the gauge variation $\delta \Psi = Q \Lambda$
in the small Hilbert space.

To summarize, the residual gauge symmetry after the partial gauge fixing
is generated by the variation $\delta \Phi$ satisfying $\Xil \delta \Phi = 0$,
and it can be identified with the gauge symmetry in the small Hilbert space
generated by the variation $\delta \Psi = Q \Lambda$.
The relation between $\LQ$ and $\Lambda$ is given by
\begin{equation}
\LQ = -\Xil \Lambda \,, \qquad
\Lambda = -\ez \LQ \,.
\label{epsilon_Q-Lambda}
\end{equation}

Let us move on to the interacting theory.
We presented the gauge transformation $\delta \Phi$
in the expansion with respect to $g$ in~\eqref{expansion:delta}: 
\begin{equation} \label{GaugeVar:Phi:app} 
\delta \Phi 
= \left( Q \LQ + \ez \Leta\right)
- \frac{g}{2} \bigl[ \Phi, Q \LQ - \ez \Leta \bigr]
+ \frac{g^2}{12} \Bigl[\Phi,\bigl[\Phi, Q \LQ + \ez \Leta \bigr]\Bigr] +O(g^3)\,.
\end{equation}
As in the free theory, we impose the constraint
\begin{equation} \label{Xil:delta}
\Xil \, \delta \Phi = 0 \,.
\end{equation}
The constraint can be solved
in the expansion with respect to $g$,
and $\ez \Leta$ can be expressed in terms of $\LQ$ and $\Phi$.
Let us expand $\ez \Leta$ in $g$ as follows:
\begin{equation} \label{exp:Leta}
\ez\Leta = A^{(0)}+ g A^{(1)} + g^2 A^{(2)} + O(g^3)\,,\qquad 
\end{equation}
where
\begin{equation}
\ez A^{(k)} = 0
\end{equation}
for $k = 0, 1, 2$.
The expansion of the constraint~\eqref{Xil:delta} in $g$ gives the following set of equations:
\bs 
\begin{align}
&\Xil\Bigl( A^{(0)} + Q\LQ \Bigr) = 0\,,
\label{1stEq} \\[.5ex]
&\Xil\biggl( A^{(1)} + \frac{1}{2}\bigl[\Phi, A^{(0)}- Q\LQ\bigr]\biggr) =0\,,
\label{2ndEq} \\[.5ex]
&\Xil\biggl(
A^{(2)} + \frac{1}{2} \bigl[\Phi, A^{(1)}\bigr]
+\frac{1}{12} \Bigl[ \Phi, \bigl[\Phi, A^{(0)} + Q\LQ\bigr]\Bigr]
\biggr) =0\,.
\label{3rdEq}
\end{align}
\es
As in the free theory, we act with $\ez$ on these equations:
\bs 
\begin{align}
& \ez \Xil\Bigl( A^{(0)} + Q\LQ \Bigr)
= A^{(0)} +\ez \Xil Q\LQ = 0\,,
\\[.5ex]
& \ez \Xil\biggl( A^{(1)} + \frac{1}{2}\bigl[\Phi, A^{(0)}- Q\LQ\bigr]\biggr)
= A^{(1)} + \ez \Xil\biggl( \frac{1}{2}\bigl[\Phi, A^{(0)}- Q\LQ\bigr]\biggr) =0\,,
\\[.5ex]
\nonumber 
& \ez \Xil\biggl(
A^{(2)} + \frac{1}{2} \bigl[\Phi, A^{(1)}\bigr]
+\frac{1}{12} \Bigl[ \Phi, \bigl[\Phi, A^{(0)} + Q\LQ\bigr]\Bigr]
\biggr) \\*  
& = A^{(2)} + \ez \Xil\biggl(
\frac{1}{2} \bigl[\Phi, A^{(1)}\bigr]
+\frac{1}{12} \Bigl[ \Phi, \bigl[\Phi, A^{(0)} + Q\LQ\bigr]\Bigr]
\biggr) =0 \,,
\end{align}
\es
where we used
$\ez\Xil A^{(k)} = \bigl(1-\Xil\ez\bigr) A^{(k)} = A^{(k)}$ for each $k$.
We can then obtain $A^{(0)}$, $A^{(1)}$, and $A^{(2)}$ from the following equations:
\bs 
\begin{align}
& A^{(0)} = -\ez\Xil Q\LQ \,,
\label{a0} \\[.5ex]
& A^{(1)} =-\ez\Xil\biggl( \frac{1}{2}\bigl[\Phi, A^{(0)}- Q\LQ\bigr] \biggr) \,,
\label{a1} \\[.5ex]
& A^{(2)} = -\ez\Xil \biggl( \frac{1}{2} \bigl[\Phi, A^{(1)}\bigr]
+\frac{1}{12} \Bigl[ \Phi, \bigl[\Phi, A^{(0)} + Q\LQ\bigr]\Bigr]
\biggr) \,.
\label{aN}
\end{align}
\es
Note that we can rewrite $\delta \Phi$
in the following $\Xil$-exact form
using these equations:
\begin{align}
\delta\Phi=
\Xil\eta_0\Biggl[Q\epsilon_Q
- \frac{g}{2}\bigl[\Phi, Q\LQ-A^{(0)}\bigr] 
+g^2\biggl( \frac{1}{2} \bigl[\Phi, A^{(1)}\bigr]
+\frac{1}{12} \Bigl[ \Phi, \bigl[\Phi, A^{(0)} + Q\LQ\bigr]\Bigr]
\biggr)\Biggr] +O(g^3) \,.
\end{align}
The variation $\delta \Phi$ is now written in terms of $\LQ$ and $\Phi$,
and we introduce the variation $\delta \Psi$
of the string field $\Psi$
in the small Hilbert space
as $\delta \Phi = \Xil \delta \Psi$.
We expand
$\delta \Psi$ 
in $g$ as
\begin{equation} \label{Psi:transf}
\delta \Psi = \delta^{(1)}\Psi + g\,\delta^{(2)}\Psi 
+ g^2\,\delta^{(3)}\Psi + O(g^3),
\end{equation}
and $\delta^{(1)}\Psi$, $\delta^{(2)}\Psi$, and $\delta^{(3)}\Psi$
are given by 
\bs \label{deltaPsi}
\begin{align}
\delta^{(1)} \Psi &= -Q \ez\LQ \,,\\[1ex]
\delta^{(2)} \Psi 
&= - \frac{1}{2}\Bigl( \bigl[ \Psi, (1 +\ez\Xil)\, Q\LQ \bigr] - \bigl[ \Xil\Psi, Q \ez\LQ \bigr] \Bigr),\\[1ex]
\delta^{(3)} \Psi 
&= \frac{1}{4} \Bigl[ \Psi, \ez\Xil \bigl[ \Xil\Psi, (1 + \ez\Xil)\, Q\LQ \bigr] \Bigr] 
\nonumber \\*
&
\quad~
+\frac{1}{12} \biggl(
\Bigl[\Psi,\bigl[\Xil\Psi, \Xil\ez Q\LQ \bigr]\Bigr] 
+\Bigl[\Xil\Psi,\bigl[\Psi, \Xil\ez Q\LQ \bigr]\Bigr] 
-\Bigl[\Xil\Psi,\bigl[\Xil\Psi, Q \ez\LQ \bigr]\Bigr] 
\biggr) \,,
\end{align}
\es
where we wrote $\Phi$ as $\Xil \Psi$.
Using the relation $\LQ = -\Xil \Lambda$ in~\eqref{epsilon_Q-Lambda},
we can express
$\delta \Psi$
in terms of $\Lambda$ and $\Psi$.
The term $(1+\ez\Xil) Q\LQ$
which appeared
in~$\delta^{(2)} \Psi$ and $\delta^{(3)} \Psi$
can be transformed in the following way:
\begin{equation}
\begin{split}
(1+\ez\Xil) \, Q \LQ
& = {}-Q \Xil \Lambda -\ez \Xil Q \Xil \Lambda
= {}-Q \Xil \Lambda -\ez \Xl \Xil \Lambda
= {}-Q \Xil \Lambda -\Xl \ez \Xil \Lambda \\
& = {}-Q \Xil \Lambda -\Xl \Lambda
= -2 \Xl \Lambda +\Xil Q \Lambda \,,
\end{split}
\end{equation}
where we used $\{ Q, \Xil \} = \Xl$, $\Xil^2 = 0$,
$[ \, \ez, \Xl \, ] = 0$, and $\ez \Xil \Lambda = ( 1 -\Xil \ez ) \, \Lambda = \Lambda$.
The final expressions of
$\delta^{(1)} \Psi$, $\delta^{(2)} \Psi$, and $\delta^{(3)} \Psi$
in terms of $\Lambda$ and $\Psi$ are
\bs \label{var}
\begin{align}
\delta^{(1)} \Psi &= Q\Lambda \,,
\label{var0-Psi} \\[1ex]
\delta^{(2)} \Psi &= - \frac{1}{2}\Bigl( \bigl[ \Psi, (-2\calX_\lambda + \Xil Q)\Lambda \bigr]
+ \bigl[ \Xil\Psi, Q\Lambda \bigr] \Bigr) ,
\label{var1-Psi} \\[1ex]
\delta^{(3)}\Psi &= \frac{1}{4} \Bigl[ \Psi, \ez\Xil \bigl[ \Xil\Psi, (-2\calX_\lambda + \Xil Q)\Lambda \bigr] \Bigr] 
\nonumber \\*
&
\quad~
+\frac{1}{12} \Bigl(
\bigl[\Psi,[\Xil\Psi, \Xil Q\Lambda ]\bigr] 
+\bigl[\Xil\Psi,[\Psi, \Xil Q\Lambda ]\bigr] 
+\bigl[\Xil\Psi,[\Xil\Psi, Q \Lambda ]\bigr] 
\Bigr).
\label{var2-Psi}
\end{align}
\es
Note that $\delta^{(2)} \Psi$ in~\eqref{delta^(2)-Psi} is reproduced
based on the residual gauge transformation
of the Berkovits formulation.

In this way, we can systematically construct the gauge transformation
in the expansion with respect to $g$.\footnote{ 
For the complete form of the residual gauge transformation, see~\cite{IT}. 
}
For the purpose of constructing the gauge transformation
in the formulation based on the small Hilbert space,
it is fine to restrict the gauge parameter $\LQ$ in the large Hilbert space
as $\LQ = -\Xil \Lambda$.
In the interacting theory, however, the $\ez$-exact part of $\LQ$
does not decouple after the partial gauge fixing,
and we may wonder what kind of gauge transformations it generates.
Actually, the gauge parameters $\LQ$ and $\Leta$ are not completely independent.
It is easily seen in the gauge transformation~\eqref{gauge-transformation-free}
of the free theory.
The transformation $Q \LQ$ with $\LQ = \ez \tilde{\epsilon}$
can also be written as the transformation $\ez \Leta$ with $\Leta = -Q \tilde{\epsilon}$.
Therefore, even if we drop the $\ez$-exact part of $\LQ$,
the whole gauge transformations can be generated
as long as we keep the whole region of $\Leta$.
The situation is more complicated in the interacting theory,
but the conclusion is the same
and no new gauge transformations are generated from
the $\ez$-exact part of $\LQ$.
See~\cite{IT} for details.
This is part of the reducibility structure in the Berkovits formulation,
and it is the source of the complication
in the gauge fixing in terms of the Batalin-Vilkovisky formalism~\cite{Berkovits:2012}.

Now that we have $\delta^{(3)} \Psi$,
it follows from the gauge invariance in the Berkovits formulation that
\begin{equation} \label{varS}
\delta^{(3)} \SB_2 + \delta^{(2)} \SB_3 + \delta^{(1)} \SB_4 = 0
\end{equation}
with $\SB_4$ given in~\eqref{quartic-regularized}.
The variation
$\delta^{(2)} \SB_3$ 
is nonvanishing
and cannot be absorbed by correcting the gauge transformation,
but the gauge invariance at $O(g^2)$ is recovered
by adding the quartic interaction $\SB_4$ in~\eqref{quartic-regularized}
to the action
with the correction term
$\delta^{(3)} \Psi$ 
given in~\eqref{var2-Psi}
to the gauge transformation.
This is the answer from the Berkovits formulation
to the question in the preceding subsection.

The part of the variation
$\delta^{(2)} \SB_3$
that is not absorbed by correcting the gauge transformation
is canceled by
$\delta^{(1)} \SB_4$. 
Let us look at the structure of this term when $\lambda$ is small.
The quartic interaction before the partial gauge fixing is given
in~\eqref{quartic-useful}.
As we explained in~\eqref{quartic-small-lambda},
the term containing $Q \ez \Phi$ gives a vanishing contribution
after the partial gauge fixing
in the limit $\lambda \to 0$
because $Q \ez \Phi$ becomes $Q \Psi$ under the partial gauge fixing
and there are no picture-changing operators localized near the midpoint.
Let us express this as
\begin{equation}
\begin{split}
\SB_4 & = \frac{\iu}{8} \,
\bigl\langle \, \Phi^2, \left(Q\Phi\right)\left(\eta_0\Phi\right) \, \bigr\rangle
-\frac{\iu}{8} \,
\bigl\langle \, \Phi^2, \left(\eta_0\Phi\right)\left(Q\Phi\right) \, \bigr\rangle
+\frac{\iu}{12} \,
\bigl\langle \, \Phi^3,  Q\eta_0\Phi \, \bigr\rangle \\
& \simeq \frac{\iu}{8} \,
\bigl\langle \, \Phi^2, \left(Q\Phi\right)\left(\eta_0\Phi\right) \, \bigr\rangle
-\frac{\iu}{8} \,
\bigl\langle \, \Phi^2, \left(\eta_0\Phi\right)\left(Q\Phi\right) \, \bigr\rangle \,.
\end{split}
\end{equation}
While we are considering the gauge variation
$\delta^{(1)} \SB_4$ 
of the quartic interaction,
the situation is similar to the calculation of on-shell four-point amplitudes
in subsection~\ref{subsec:on-shell_berkovits}
because we can effectively drop terms containing $Q \ez \Phi$.
Then the variation of $\SB_4$ can be evaluated
by a calculation similar to the one
we did in deriving~\eqref{A_4-simplified}. We find
\begin{align}
\nonumber
& \delta^{(1)} \, \bigl\langle \, \Phi^2, ( Q \Phi ) ( \ez \Phi ) \, \bigr\rangle
-\delta^{(1)} \, \bigl\langle \, \Phi^2, ( \ez \Phi ) ( Q \Phi ) \, \bigr\rangle \\
\nonumber
& =
\bigl\langle \, Q \Phi, ( \ez \delta^{(1)} \Phi ) \, \Phi^2 +\delta^{(1)} \Phi \, \Phi \, ( \ez \Phi ) \, \bigr\rangle
+\bigl\langle \, Q \Phi, \Phi^2 \, ( \ez \delta^{(1)} \Phi ) +( \ez \Phi ) \, \Phi \, \delta^{(1)} \Phi \, \bigr\rangle \\
\nonumber
& \quad ~
-\bigl\langle \, \ez \Phi, ( Q \delta^{(1)} \Phi ) \, \Phi^2 +\delta^{(1)} \Phi \, \Phi \, ( Q \Phi ) \, \bigr\rangle
-\bigl\langle \, \ez \Phi, \Phi^2 \, ( Q \delta^{(1)} \Phi ) +( Q \Phi ) \, \Phi \, \delta^{(1)} \Phi \, \bigr\rangle \\
\nonumber
& \simeq
-\bigl\langle \, Q \Phi, \delta^{(1)} \Phi \, ( \ez \Phi ) \, \Phi \, \bigr\rangle
-\bigl\langle \, Q \Phi, \Phi \, ( \ez \Phi ) \, \delta^{(1)} \Phi \, \bigr\rangle
+\bigl\langle \, \ez \Phi, \delta^{(1)} \Phi \, ( Q \Phi ) \, \Phi \, \bigr\rangle
+\bigl\langle \, \ez \Phi, \Phi \, ( Q \Phi ) \, \delta^{(1)} \Phi \, \bigr\rangle \\
& = 2 \, \bigl\langle \, ( \ez \Phi ) \, \Phi \, ( Q \Phi ), \delta^{(1)} \Phi \, \bigr\rangle
-2 \, \bigl\langle \, ( Q \Phi ) \, \Phi \, ( \ez \Phi ), \delta^{(1)} \Phi \, \bigr\rangle \,,
\end{align}
where we dropped terms containing $Q \ez \Phi$.
Therefore,
$\delta^{(1)} \SB_4$
for small $\lambda$
is given by
\begin{equation}
\delta^{(1)} \SB_4
\simeq \frac{\iu}{4} \, \biggl[ \,
\Big\langle \Psi \bigl( \Xil \Psi \bigr) \bigl( \Xl \Psi \bigr), \bigl( \Xil Q \Lambda \bigr) \Big\rangle
-\Big\langle \bigl( \Xl \Psi \bigr) \bigl( \Xil \Psi \bigr) \Psi, \bigl( \Xil Q \Lambda \bigr) \Big\rangle
\, \biggr] \,.
\label{delta-S_4} 
\end{equation} 
As we explained in subsection~\ref{section:vertices},
$\Xl$ and $\Xil$ for small $\lambda$ are approximated
by insertions of local operators $X$ and $\xi$, respectively,
near the midpoint.
In either of the two terms on the right-hand side of~\eqref{delta-S_4},
we therefore have two insertions of $\xi$ and one insertion of $X$
near the midpoint.
While the positions of two insertions of $\xi$ are the same
in both terms,
the position of the picture-changing operator $X$ is different.
It approaches the midpoint in two different ways,
and
$\delta^{(1)} \SB_4$
is nonvanishing
and divergent in the limit $\lambda \to 0$.
This can be explicitly confirmed by using the OPE of $X$ and $\xi$.

To summarize,
the variation of the cubic interaction
$\delta^{(2)} \SB_3$, 
which is not well defined in the Witten formulation,
is regularized in our approach,
and the leading behavior in the limit $\lambda \to 0$
of the part of~$\delta^{(2)} \SB_3$ 
that is not absorbed by correcting the gauge transformation
takes the form of the difference between two terms
where the picture-changing operator approaches the midpoint
in different ways.
The gauge invariance at $O(g^2)$ is recovered
by incorporating the quartic interaction~\eqref{quartic-regularized}
and its role is to adjust the behavior of the picture-changing operator,
which is analogous to the picture we found
in our analysis of on-shell four-point amplitudes in section~\ref{section:regularization}.

\section{Discussion}
\label{section:summary}
\setcounter{equation}{0}
Our primary motivation for the condition 
$\Xil \Phi = 0$ in \eqref{gauge-condition}
for the partial gauge fixing
is to discuss the relation between the Berkovits formulation
and the Witten formulation.
In our discussion of the on-shell four-point amplitude in section~\ref{section:regularization},
we learned that the quartic interaction plays a role
of adjusting different behaviors of the picture-changing operators
in the $s$ channel and in the $t$ channel
of Feynman diagrams with two cubic vertices.
While we considered the limit $\lambda \to 0$ in subsection~\ref{Relation_to_Witten}
to discuss the nature of the divergence in the Witten formulation,
our discussion in subsection~\ref{subsec:on-shell_berkovits}
is general and does not depend on the limit $\lambda \to 0$.
In fact, we can make it even more general
by replacing $\Xil$ with a more general line integral of $\xi$,
which we denote by $\Xi$ as in subsection~\ref{subsec:idea}.

In our discussion of the gauge invariance at $O(g^2)$ in section~\ref{sec:relation},
we learned that the gauge variation $\delta^{(2)} \SB_3$ of the cubic interaction
is not completely absorbed by correcting the gauge transformation
and the gauge invariance is recovered by incorporating the quartic interaction.
While we considered the limit $\lambda \to 0$
to discuss the relation to the divergence in the Witten formulation,
the theory constructed by the partial gauge fixing
is gauge invariant for any $\lambda$.
In fact, we can construct a more general class of gauge-invariant theories
by replacing $\Xil$ with $\Xi$.

One important lesson from the discussions in this paper is that
we have actually succeeded in constructing a consistent formulation
of open superstring field theory based on the small Hilbert space
by partial gauge fixing of the Berkovits formulation.
The complication of open superstring field theory based on the small Hilbert space
is usually associated with the necessity of local picture-changing operators,
but the theory obtained by the partial gauge fixing uses a line integral $\Xi$
and is free from singularity coming from local picture-changing operators.
We can, for example, choose $\Xi$ to be $\xi_0$. 

Apparently, it is surprising that the theory with a general line integral $\Xi$
is gauge invariant.
Let us compare $\Xi$ with the BRST operator $Q$.
The BRST current is a primary field of conformal weight~$1$,
and thus the zero mode $Q$
acts as a derivation
with respect to the star product:
\begin{equation}
Q \, ( A \ast B ) = Q A \ast B +(-1)^A A \ast Q B
\end{equation}
for any states $A$ and $B$.
The derivation property of $Q$ is crucial
for gauge invariance in open string field theory
with string products defined by the star product.
Similarly, $\ez$ acts as a derivation
with respect to the star product,
and this property is also crucial for gauge invariance
in the Berkovits formulation of open superstring field theory.
On the other hand, $X$ and $\xi$ are primary fields of conformal weight 
$0$,
so their zero modes or general line integrals do not have
simple transformation properties under the conformal map
associated with the star product.
For example, the state $\Xi \, ( A \ast B )$ does not have
any simple relations to $\Xi A \ast B$ and $A \ast \Xi B$ in general.
Local operators of $X$ and $\xi$ have simple transformation properties
under conformal maps, but they have to be inserted
at the midpoint if we want the properties~\eqref{X_mid-identities}
or~\eqref{xi_mid-identities} to hold.
The reason why the theory with $\Xi$ is gauge invariant
is that we never need to deform the contour
of the line integral $\Xi$ in proving the gauge invariance.
This observation will open up new possibilities
for string field theory based on the small Hilbert space.
The fact that we can construct a consistent theory
based on the small Hilbert space
using a class of general operators $\Xi$
should have an implication
in the context of the covering of the supermoduli space
of super-Riemann surfaces.

While the gauge invariance of the theory obtained by the partial gauge fixing
does not depend on the form of $\Xi$,
the combinatorial aspect of the quartic interaction
inherited from the Berkovits formulation was crucial
and it should also have an implication
in the context of the covering of the supermoduli space
of super-Riemann surfaces.
The combinatorial aspect of interaction vertices
is in a sense obscured in the beautiful WZW-like form
of the action in the Berkovits formulation,
and it would be important to decode it.
Extension of the discussion in subsection~\ref{subsec:on-shell_berkovits}
to higher-point amplitudes would be useful
and will be discussed in~\cite{GIN}.
Extension to off-shell amplitudes will also be discussed in~\cite{GIN}.

Another important direction is to extend the discussion
of the relation between the Berkovits formulation
and the Witten formulation
in the framework of the Batalin-Vilkovisky formalism.
The correspondence of the kinetic term
of the master action 
in the Batalin-Vilkovisky formalism
between the Berkovits formulation and the Witten formulation
was discussed in~\cite{Kroyter:2012ni}.
As we mentioned in the introduction,
the construction of the master action
for the interacting theory
has turned out to be complicated for the Berkovits formulation,
but the complete form of cubic terms
is derived~\cite{Torii:proceedings, Berkovits:2012}.
In fact, by extending the one-parameter family of gauges \eqref{gauge-condition} to the sector of ghost string fields 
in the master action,
it can be shown
that these cubic interactions of the master action correspond to those of the Witten formulation
in the singular limit $\lambda\to 0$~\cite{IT}.
In general, the form of the master action is governed by the reducibility structure,
which is the gauge structure more detailed than that investigated in section~\ref{sec:relation}.
The relation between the reducibility structures of the two formulations is also elucidated in \cite{IT}.

\section*{Acknowledgments}
We would like to thank Keiyu Goto and Koichi Murakami for useful discussions.
The work of Y.I.\ was
supported in part
by the Grand-in-Aid for
Nagoya University Leadership Development Program for Space Exploration and Research.
The work of T.N.\ and of S.T.\ were supported in part by the Special Postdoctoral Researcher Program at RIKEN.
The work of Y.O. was supported in part
by Grant-in-Aid for Scientific Research~(B) No.~25287049
and Grant-in-Aid for Scientific Research~(C) No.~24540254
from the Japan Society for the Promotion of Science (JSPS).
The work of Y.O. and S.T. was also supported in part
by the M\v{S}MT contract No. LH11106 and
by JSPS and the Academy of Sciences of the Czech Republic (ASCR)
under the Research Cooperative Program between Japan and the Czech Republic.

\appendix

\section{The partial gauge fixing and the reality condition}
\label{section:reality}
\setcounter{equation}{0}
The reality condition on the string field guarantees
that the value of the action is real~\cite{Gaberdiel:1997ia}.
In the Berkovits formulation of open superstring field theory,
the reality condition on the superstring field $\Phi$ in the NS sector
can be stated as
\begin{equation}
\Phi^\dagger = -\Phi^\star \,.
\label{reality-Phi}
\end{equation}
We denote the Hermitian conjugate of $A$ by $A^\dagger$
and the BPZ conjugate of $A$ by $A^\star$,
where $A$ can be a string field or an operator.
For detailed discussion on the condition~(\ref{reality-Phi}),
see appendix B of~\cite{Torii:validity}.

In our discussion of the partial gauge fixing, we consider the condition
\begin{equation}
\Xi \Phi = 0 \,,
\label{gauge-condition-appendix}
\end{equation}
where the Grassmann-odd operator $\Xi$ satisfies $\Xi^2 = 0$ and $\{ \Xi, \ez \} = 1$.
The ghost number of $\Xi$ is $-1$ and the picture number of $\Xi$ is $1$.
For the partial gauge fixing to be compatible with the reality condition on $\Phi$,
the condition~(\ref{gauge-condition-appendix}) has to be solved
by a string field satisfying the reality condition.
We know from the discussion in section~\ref{section:Gauge-fixing} that
the solution to the condition~(\ref{gauge-condition-appendix}) is given by $\Xi \ez \Phi$.
Therefore, the condition~(\ref{gauge-condition-appendix}) is compatible
with the reality condition~(\ref{reality-Phi}) if
\begin{equation}
( \, \Xi \ez \Phi \, )^\dagger = -( \, \Xi \ez \Phi \, )^\star \,.
\end{equation}
It is convenient to use the standard bra-ket notation for the Hermitian conjugation.
We write the Hermitian conjugate of $| \Phi \rangle$ as $\langle \Phi |$,
which should not be confused with the BPZ conjugate.
We have
\begin{equation}
( \, \Xi \ez | \Phi \rangle \, )^\dagger
= \langle \Phi | \, \ez^\dagger \, \Xi^\dagger
= \langle \Phi | \, \ez\, \Xi^\dagger \,. 
\end{equation}
On the other hand, the BPZ conjugate of $\Xi \ez | \Phi \rangle$ is
\begin{equation}
( \, \Xi \ez | \Phi \rangle \, )^\star
= (-1)^2 \langle \Phi | \, \ez^\star \, \Xi^\star
= -\langle \Phi | \, \ez\, \Xi^\star \,,
\end{equation}
where one minus sign is from the reality condition~(\ref{reality-Phi}) on $\Phi$
and the other is from the definition of the BPZ conjugation
when both $\Xi$ and $\ez$ are Grassmann odd and $\Phi$ is Grassmann even.
We thus conclude that the condition~(\ref{gauge-condition-appendix}) is compatible
with the reality condition~(\ref{reality-Phi}) if
\begin{equation}
\Xi^\dagger = \Xi^\star \,.
\label{reality-Xi}
\end{equation}

It is straightforward to verify that the operator $\Xil$
with the expansion~(\ref{Xil-expansion}) satisfies the condition~(\ref{reality-Xi}).
Since $\xi_n^\dagger = \xi_{-n}$, we find that $\Xil^\dagger = \Xil$.
As we already mentioned in section~\ref{section:Gauge-fixing},
$\Xil$ is BPZ even because $\xi_n^\star = (-1)^n \, \xi_{-n}$.
We thus have $\Xil^\dagger = \Xil^\star$
and we conclude that the partial gauge fixing
with the condition $\Xil \Phi = 0$
is compatible with the reality condition~(\ref{reality-Phi}) on $\Phi$.

\section{Modification of picture-changing operators}
\label{subsub:more_general}
\setcounter{equation}{0}
In subsection~\ref{subsec:super_decomposition},
the color-ordered amplitude $\mathcal{A}^{\rm ws}_{ABCD}$ in the world-sheet theory
was written in the following form in~\eqref{WS_SFT-like}:
\begin{equation}
\mathcal{A}^{\rm ws}_{ABCD}
= g^2 \, \Bllangle X_0\Psi_A\ast X_0\Psi_B\,,
\frac{b_0}{L_0} \,
(\Psi_C\ast \Psi_D)\Brrangle
+g^2 \, \Bllangle  X_0\Psi_B\ast \Psi_C\,,
\frac{b_0}{L_0} \,
(\Psi_D\ast X_0\Psi_A)\Brrangle\,.
\label{A^ws_ABCD-1}
\end{equation}
In this appendix, we first show that the locations of the picture-changing operators
can be changed.
We then show that the operator $X_0$ can be replaced with $\Xl$ defined in~\eqref{calX}.

In~\eqref{A^ws_ABCD-1}, the picture-changing operators are acting
on the states $\Psi_A$ and $\Psi_B$.
Consider moving $X_0$ on $\Psi_B$ to $\Psi_C$. The procedure is as follows.
First, write $X_0$ in front of $\Psi_B$ as $\{ Q, \xi_0 \}$.
Second, add $\xi_0$ in front of $\Psi_C$
to express the BPZ inner product in the large Hilbert space.
Third, move $Q$ to act on $\xi_0$ and $b_0 / L_0$.
For the $s$-channel contribution, we have
\begin{align}
\nonumber
&\Bllangle X_0 \Psi_A\ast X_0 \Psi_B\,,
\frac{b_0}{L_0} \,
(\Psi_C\ast \Psi_D)\Brrangle\\*
\nonumber
&=-\iu\,\Big\langle X_0 \Psi_A\ast \{Q,\xi_0\}\Psi_B\,,
\frac{b_0}{L_0} \,
(\xi_0 \Psi_C\ast \Psi_D)\Big\rangle\\*
\nonumber
&=-\iu\,\Big\langle X_0 \Psi_A\ast \xi_0 \Psi_B\,,
\frac{b_0}{L_0} \,
( X_0 \Psi_C\ast \Psi_D)\Big\rangle
+\iu\,\Big\langle X_0 \Psi_A\ast \xi_0 \Psi_B\,,
\xi_0 \Psi_C\ast \Psi_D\Big\rangle\\*
&=\Bllangle X_0 \Psi_A\ast \Psi_B\,,
\frac{b_0}{L_0} \,
( X_0 \Psi_C\ast \Psi_D)\Brrangle
+\iu\,\Big\langle X_0 \Psi_A\ast \xi_0 \Psi_B\,,
\xi_0 \Psi_C\ast \Psi_D\Big\rangle\,.
\label{B-to-C-s}
\end{align}
The second term in the last line can be interpreted
as a contribution from the boundary of the moduli space
in the $s$ channel.
For the $t$-channel contribution, we have
\begin{align}
\nonumber
&\Bllangle X_0 \Psi_B\ast \Psi_C\,,
\frac{b_0}{L_0} \,
(\Psi_D\ast X_0 \Psi_A)\Brrangle\\*
\nonumber
&=\Bllangle  \Psi_B\ast X_0 \Psi_C\,,
\frac{b_0}{L_0} \,
(\Psi_D\ast X_0 \Psi_A)\Brrangle
+\iu \, \Big\langle  \xi_0 \Psi_B\ast \xi_0\Psi_C\,, 
\Psi_D\ast X_0 \Psi_A\Big\rangle\\*
&=\Bllangle  \Psi_B\ast X_0 \Psi_C\,,
\frac{b_0}{L_0} \,
(\Psi_D\ast X_0 \Psi_A)\Brrangle
-\iu \, \Big\langle X_0 \Psi_A\ast \xi_0 \Psi_B\,, 
\xi_0 \Psi_C\ast\Psi_D\Big\rangle\,,
\label{B-to-C-t}
\end{align}
where we used the cyclicity property of BPZ inner products
consisting of star products of string fields.
We find that
the boundary term in the $s$ channel
and the boundary term in the $t$ channel cancel,
and we obtain the expression
where the picture-changing operators on $\Psi_B$
have been moved to $\Psi_C$ both
in the $s$ channel and in the $t$ channel:
\begin{equation}
\begin{split}
& \Bllangle X_0\Psi_A\ast X_0\Psi_B\,,
\frac{b_0}{L_0} \,
(\Psi_C\ast \Psi_D)\Brrangle
+\Bllangle  X_0\Psi_B\ast \Psi_C\,,
\frac{b_0}{L_0} \,
(\Psi_D\ast X_0\Psi_A)\Brrangle \\
& = \Bllangle X_0\Psi_A\ast \Psi_B\,,
\frac{b_0}{L_0} \,
( X_0 \Psi_C\ast \Psi_D)\Brrangle
+\Bllangle  \Psi_B\ast X_0 \Psi_C\,,
\frac{b_0}{L_0} \,
(\Psi_D\ast X_0\Psi_A)\Brrangle \,.
\end{split}
\end{equation}
It is clear that we can move $X_0$ from any state to any state
as long as we do the same move in the $s$ channel and in the $t$ channel.

We next show that $X_0$ in~\eqref{A^ws_ABCD-1} can be replaced with $\Xl$.
The operator $X_0$ is $\Xl$
at $\lambda = \infty$,\footnote{
Note that this limit is regular in contrast to
the 
limit $\lambda\to0$.}
and we actually show a more general formula given by
\begin{align}
\nonumber
& \Bllangle \mathcal{X}_{\lambda_1}\Psi_A\ast \mathcal{X}_{\lambda_3}\Psi_B\,,
\frac{b_0}{L_0} \,
(\Psi_C\ast \Psi_D)\Brrangle
+\Bllangle  \mathcal{X}_{\lambda_3}\Psi_B\ast \Psi_C\,,
\frac{b_0}{L_0} \,
(\Psi_D\ast \mathcal{X}_{\lambda_1}\Psi_A)\Brrangle\\*
&= \Bllangle\mathcal{X}_{\lambda_2} \Psi_A\ast \mathcal{X}_{\lambda_3}\Psi_B\,,
\frac{b_0}{L_0} \,
(\Psi_C\ast \Psi_D)\Brrangle
+\Bllangle  \mathcal{X}_{\lambda_3}\Psi_B\ast \Psi_C\,,
\frac{b_0}{L_0} \,
(\Psi_D\ast \mathcal{X}_{\lambda_2}\Psi_A)\Brrangle
\label{general-lambda}
\end{align}
for any $\lambda_1$, $\lambda_2$, and $\lambda_3$.
The $s$-channel contribution in the first line can be transformed as follows:
\begin{align}
\nonumber
&\Bllangle \mathcal{X}_{\lambda_1}\Psi_A\ast \mathcal{X}_{\lambda_3}\Psi_B,
\frac{b_0}{L_0} \,
(\Psi_C\ast \Psi_D)\Brrangle
\\*
\nonumber
&=
\Bllangle \mathcal{X}_{\lambda_2}\Psi_A\ast \mathcal{X}_{\lambda_3}\Psi_B,
\frac{b_0}{L_0} \,
(\Psi_C\ast \Psi_D)\Brrangle
+
\Bllangle ( \mathcal{X}_{\lambda_1}-\mathcal{X}_{\lambda_2} ) \, \Psi_A
\ast \mathcal{X}_{\lambda_3}\Psi_B,
\frac{b_0}{L_0} \,
\left(\Psi_C\ast \Psi_D\right)\Brrangle
\\*
\nonumber
&=
\Bllangle \mathcal{X}_{\lambda_2}\Psi_A\ast \mathcal{X}_{\lambda_3}\Psi_B,
\frac{b_0}{L_0} \,
(\Psi_C\ast \Psi_D)\Brrangle
+
\Bllangle \{Q,\, \Xi_{\lambda_1}-\Xi_{\lambda_2}\} \, \Psi_A
\ast \mathcal{X}_{\lambda_3}\Psi_B,
\frac{b_0}{L_0} \,
(\Psi_C\ast \Psi_D)\Brrangle
\\*
&=
\Bllangle \mathcal{X}_{\lambda_2}\Psi_A\ast \mathcal{X}_{\lambda_3}\Psi_B,
\frac{b_0}{L_0} \,
(\Psi_C\ast \Psi_D)\Brrangle
+
\Bllangle ( \, \Xi_{\lambda_1}-\Xi_{\lambda_2} ) \, \Psi_A \ast \mathcal{X}_{\lambda_3}\Psi_B,
\Psi_C\ast \Psi_D\Brrangle\,.
\label{general-lambda-s}
\end{align}
The second term in the last line can be again interpreted
as a contribution from the boundary of the moduli space
in the $s$ channel.
Here it is important that all the states are in the small Hilbert space.
Note, in particular, that the state $( \, \Xi_{\lambda_1}-\Xi_{\lambda_2} ) \, \Psi_A$
is in the small Hilbert space because
$\{ \, \ez,\, \Xi_{\lambda_1}-\Xi_{\lambda_2} \} = 0$.
Similarly, the $t$-channel contribution can be transformed as follows:
\begin{align}
\nonumber
&\Bllangle  \mathcal{X}_{\lambda_3}\Psi_B\ast \Psi_C\,,
\frac{b_0}{L_0} \,
(\Psi_D\ast \mathcal{X}_{\lambda_1}\Psi_A)\Brrangle\\*
\nonumber
&=
\Bllangle  \mathcal{X}_{\lambda_3}\Psi_B\ast \Psi_C\,,
\frac{b_0}{L_0} \,
(\Psi_D\ast \mathcal{X}_{\lambda_2}\Psi_A)\Brrangle
-\Bllangle  \mathcal{X}_{\lambda_3}\Psi_B\ast \Psi_C\,,\Psi_D
\ast ( \, \Xi_{\lambda_1}-\Xi_{\lambda_2} ) \,\Psi_A\Brrangle
\\*
&=
\Bllangle  \mathcal{X}_{\lambda_3}\Psi_B\ast \Psi_C\,,
\frac{b_0}{L_0} \,
(\Psi_D\ast \mathcal{X}_{\lambda_2} \Psi_A)\Brrangle
-\Bllangle ( \, \Xi_{\lambda_1}-\Xi_{\lambda_2} ) \,\Psi_A
\ast \mathcal{X}_{\lambda_3}\Psi_B \,,\Psi_C\ast\Psi_D\Brrangle \,, 
\label{general-lambda-t}
\end{align}
where we used the cyclicity property of BPZ inner products
consisting of star products of string fields.
We find that
the boundary term in the $s$ channel
and the boundary term in the $t$ channel cancel,
and we obtain the formula~\eqref{general-lambda}.
The picture-changing operators are acting on $\Psi_A$ and $\Psi_B$
in~\eqref{general-lambda},
but it is clear that this formula can be generalized
to the case where two picture-changing operators are acting on any states
as long as the two operators are acting on the same states
in the $s$ channel and in the $t$ channel.
It is also clear that we can change the value of $\lambda$ for any $\Xl$
as long as we do the same change in the $s$ channel and in the $t$ channel.


\end{document}